\documentclass[fleqn,10pt]{wlscirep}
\usepackage[utf8]{inputenc}
\usepackage[T1]{fontenc}
\usepackage{graphicx,wrapfig}
\usepackage{dcolumn}
\usepackage{bm}
\usepackage{times}
\usepackage{subcaption,bbm}
\usepackage{physics}
\allowdisplaybreaks
\usepackage{colortbl}
\usepackage{tikz}
\usetikzlibrary{arrows.meta}
\usepackage{hyperref}
\usepackage{cases}
\usepackage{multirow}
\usepackage{mathtools}
\usepackage{comment}

\usepackage{caption}
\newcommand{\ii}{\mathrm{i}}
\newcommand{\ee}{\mathrm{e}}
\newcommand{\st}{\mathrm{st}}

\newcommand{\lk}{\ket{\mathrm{L}}}

\newcommand{\rk}{\ket{\mathrm{R}}}

\newcommand{\gs}{\ket{\mathrm{G}}}
\newcommand{\es}{\ket{\mathrm{E}}}
\newcommand{\psiR}{\psi^{\mathrm{R}}}
\newcommand{\psiL}{\psi^{\mathrm{L}}}
\newcommand{\NH}{\mathrm{NH}}
\newcommand{\ctext}[1]{\raise0.2ex\hbox{\textcircled{\scriptsize{#1}}}}

\title{Proposal of a quantum version of active particles via a nonunitary quantum walk}

\author[1,2,*]{Manami Yamagishi}
\author[3]{Naomichi Hatano}
\author[4,3]{Hideaki Obuse}
\affil[1]{The University of Tokyo, Department of Physics, Chiba, 277-8574, Japan}
\affil[2]{RIKEN, Theoretical Quantum Physics Laboratory, Saitama, 351-0198, Japan}
\affil[3]{The University of Tokyo, Institute of Industrial Science, Chiba, 277-8574, Japan}
\affil[4]{Hokkaido University, Department of Applied Physics, Hokkaido, 060-8628, Japan}

\affil[*]{manami@iis.u-tokyo.ac.jp}

\begin{abstract}
The main aim of the present paper is to define an active particle in a quantum framework as a minimal model of quantum active matter and investigate the differences and similarities of quantum and classical active matter.
Although the field of active matter has been expanding, most research has been conducted on classical systems.
Here, we propose a truly deterministic quantum active-particle model with a nonunitary quantum walk as the minimal model of quantum active matter.
We aim to reproduce results obtained previously with classical active Brownian particles; that is, a Brownian particle, with finite energy take-up, becomes active and climbs up a potential wall.
We realize such a system with nonunitary quantum walks.
We introduce new internal states, the ground state $\gs$ and the excited state $\es$, and a new nonunitary operator $N(g)$ for an asymmetric transition between $\gs$ and $\es$.
The non-Hermiticity parameter $g$ promotes the transition to the excited state; hence, the particle takes up energy from the environment.
For our quantum active particle, we successfully observe that the movement of the quantum walker becomes more active in a nontrivial manner as we increase the non-Hermiticity parameter $g$, which is similar to the classical active Brownian particle.
We also observe three unique features of quantum walks, namely, ballistic propagation of peaks in one dimension, the walker staying on the constant energy plane in two dimensions, and oscillations originating from the resonant transition between the ground state $\gs$ and the excited state $\es$ both in one and two dimensions.
\end{abstract}

\begin{document}

\flushbottom
\maketitle

%
%

\thispagestyle{empty}


\section*{Introduction}

\begin{figure}
    \centering
      \includegraphics[width=0.6\columnwidth]{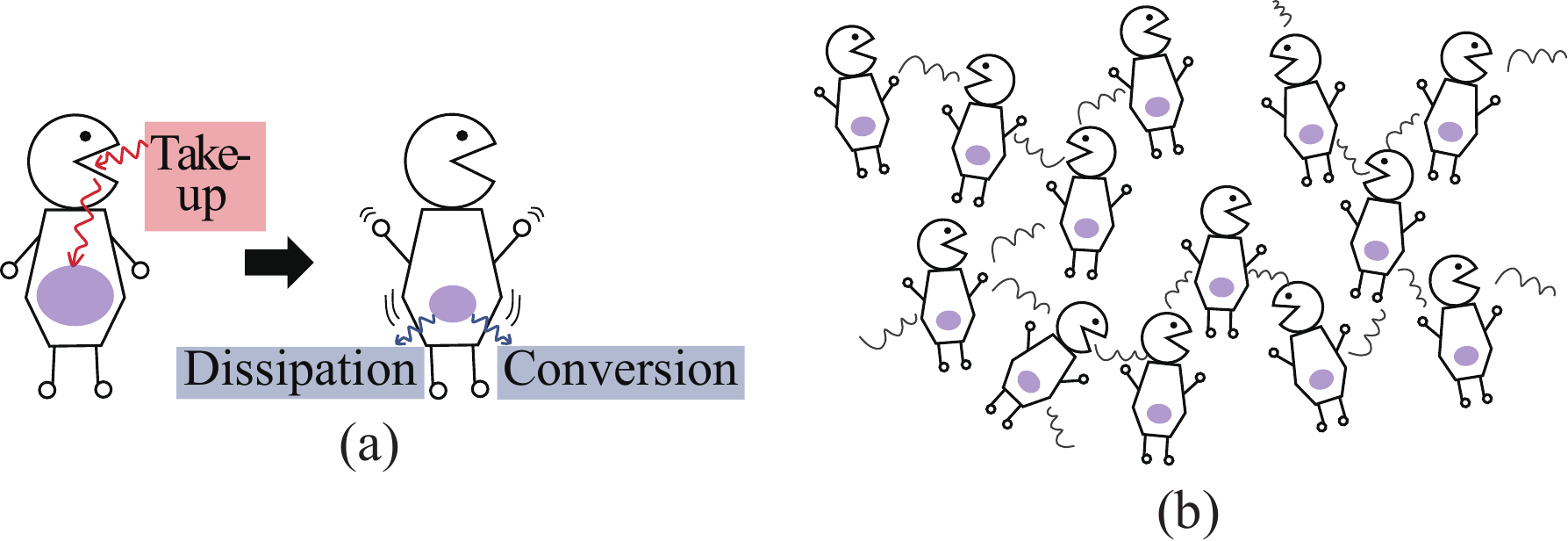}
  \caption{Schematic views of active matter. (a) A component that takes up energy from the environment, stores it internally, converts the stored internal energy to the kinetic energy and moves, and (b) an interacting collection of such components.}
  \label{packmen}
\end{figure}

Active matter is a self-driven component or a collection of such components~\cite{Pismen21}.
Active matter can include lifeless matter as well as living matter such as birds and fish.
From a physical point of view, active matter takes energy from the environment and stores it as internal energy, converts the internal energy into kinetic energy, and thereby moves (Fig.~\ref{packmen}).
The active Brownian particle~\cite{Romanczuk12}, which appears in the following sections, is a prototypical example of active matter.

Research on active matter is highly interdisciplinary, encompassing biology~\cite{Popkin16,Needleman17}, chemistry~\cite{Doostmohammadi18,Debnath19}, and physics~\cite{Gompper20}.
The introduction of the idea of active matter has enabled the unification of a variety of studies outside conventional classes of dynamics~\cite{Hohenberg-Halperin77} that had been previously investigated separately, leading to the understanding of their common features and universal behaviors~\cite{Ohta15}.
Starting from the models proposed separately by Vicsek~\cite{Vicsek95} and Toner and Tu~\cite{Toner95} in 1995, theoretical studies have lead the research of classical active matter.
Various phenomena unique to active matter have been found, \textit{e.g.}\ true long-range order~\cite{Vicsek95}, giant number fluctuation~\cite{Toner98, Chate08} and motility-induced phase separation~\cite{Cates13, Cates15}.
Recent research~\cite{Sone20} connecting classical active matter and topological phenomena, such as the Hall effect, has been attracting much attention.
These studies have attracted and fascinated many researchers, and now that an increasing number of models have been realized in experiments~\cite{Narayan07, Zhang10, Redner13, Kawaguchi17, Nishiguchi18}, leading to an ongoing broadening of the scope of this research field.

However, most research has been conducted on classical systems.
Similarly to the research on classical active matter, the investigations of systems from the different perspective of quantum active matter will open a new research direction.
Very recently, some researchers have sought to introduce the concept of active matter into quantum systems.
These efforts followed two main approaches; let us call them the quantum-classical correspondence approach and the bottom-up approach.
In the former approach, quantum systems for which obvious classical limits to specific classical active matter models can be taken are studied.
There are three works in this category.
Adachi \textit{et al.}~\cite{Adachi22} modeled a many-body version of quantum active matter, connecting a classical stochastic active matter  to a non-Hermitian quantum spin systems, which they referred to as a ``stoquastic'' Hamiltonian.
The authors recently considered another one-dimensional model~\cite{Takasan24} on the basis of their first model~\cite{Adachi22}, finding a flocking phase equivalent in the ferromagnetism behavior of a quantum spin model.
Recently, another work on active quantum flocks~\cite{Khasseh23} has been reported by Khasseh \textit{et al}.
For the bottom-up approach, on the other hand, people come up with a  one-body quantum active particles constitute a minimal model of many-body quantum active matter.
There have been two works in this category.
The present paper is the first.
The second is the work of Zheng \textit{et al.}
~\cite{Zheng23} who used a quantum harmonic oscillator with its potential minima externally driven by stochastic active dynamics.

We note here the fact that although the main stream of classical active matter research has focused on many-body systems as the active matter, there are several important studies on one-body systems, namely, the active particle;
for instance, an active Brownian particle (cf.~Ref.~\cite{Schweitzer98,Ghosh21,Romanczuk12,Bechinger16}) and an active Ornstein Uhlenbeck particle (cf.~Ref.~\cite{Szamel14,Martin21}) have also been reported.
Some models of the active particle constitute building blocks of many-body systems, namely, the active matter.
Our bottom-up approach is motivated by this fact; similar to studies on classical active particles, we start with a one-body system and aim to eventually study quantum active matter, introducing interactions between the particles.

\begin{table}
  \caption{Classification of systems without energy conservation. Studying quantum active matter enables us to extend non-Hermitian physics to systems without momentum conservation and to extend classical active matter to quantum systems.}
  \label{tab1}
  \begin{center}
  \begin{tikzpicture}
    \fill[orange] (6.5,0) rectangle (11,1);
    \draw[thick] (0,0)--(11,0)--(11,3)--(0,3)--(0,0);
    \draw[thick] (2,3)--(2,0);
    \draw[thick] (6.5,3)--(6.5,0);
    \draw[thick] (0,2)--(11,2);
    \draw[thick] (0,1)--(11,1);
    \draw (1,0.75)node[below]{Quantum};
    \draw (1,1.75)node[below]{Classical};
    \draw (4.25,0.75)node[below]{Non-Hermitian physics};
    \draw (4.25,1.75)node[below]{Dissipative systems};
    \draw (4.25,2.75)node[below]{Momentum conserved};
    \draw (8.75,0.75)node[below]{\textbf{Quantum active matter}};
    \draw (8.75,1.75)node[below]{Classical active matter};
    \draw (8.75,2.75)node[below]{Momentum not conserved};
    \draw[line width = 3, teal, arrows = {-Stealth[inset=0pt, angle=75:5pt, scale=1.5]}] (6.2,0.5) -- (6.8,0.5);
    \draw[line width = 3, teal, arrows = {-Stealth[inset=0pt, angle=75:5pt, scale=1.5]}] (8.75,1.3) -- (8.75,0.7);
  \end{tikzpicture}
  \end{center}
\end{table}
Let us consider a diagram shown in Table~\ref{tab1}.
We believe that the following three properties are essential for a system to be an active matter:
	\setlength{\topsep}{0pt}
\begin{enumerate}
	\setlength{\itemsep}{0pt}      
	\setlength{\parskip}{0pt}      
	\setlength{\itemindent}{10pt}   
	\setlength{\labelsep}{5pt}     
  \renewcommand{\labelenumi}{(\roman{enumi})}
  \item particles take up energy from the environment to drive themselves, and hence the energy is not conserved in general;
  \item the momentum is not conserved due to self-driving and the breaking of the law of action and reaction (not due to spatial randomness);
  \item the kinetic motion depends on the internal states of the particles. 
\end{enumerate}
Energy nonconservation results in temporally inhomogeneous dynamics, such as decay and growth, whereas momentum nonconservation, which is equivalent to the breakdown of the law of action and reaction, results in spatially inhomogeneous dynamics, such as a pair of birds meeting up and flying together.
We define distinctive symmetry classes of quantum active matter by updating non-Hermitian, energy nonconservative quantum models into the new realm of momentum nonconservation.

Let us mention the essential differences between quantum active matter systems and autonomous quantum systems.
One significant difference is nonconservation of the norm; since quantum active matter systems are open systems, their dynamics are nonunitary, resulting in nonconservation of the norm, whereas the norm is generally conserved in autonomous quantum systems, whose dynamics are unitary.
Another difference is whether systems reach steady states.
Autonomous quantum systems usually exhibit decay processes, which lead systems to steady states, and the static properties of such systems are often studied, while quantum active matter systems remain nonequilibrium; in fact, active systems should always be in nonequilibrium by definition.

In the present work, to find a minimal model of quantum active matter, we define a quantum active particle as a one-particle non-Hermitian quantum system that exhibits real-time evolution in a fully quantum range without external manipulation, as in Ref.~\cite{Zheng23}, using discrete-time quantum walks~\cite{Aharonov93,Meyer96,Farhi98,Ambainis12,Asaka21}.
Since the quantum walk does not show any stochasticity of classical random walks and does not have any classical limits, neither does our model.
In our quantum active-matter model, internal states that are strongly correlated with the environment dominate the system dynamics.
A strong correlation with the environment makes the system open and non-Hermitian with no energy conservation.

We note that the non-Hermiticity of our model Hamiltonian belongs to the class of so-called pseudo-Hermiticity~\cite{Mostafazadeh02-1,Mostafazadeh02-2,Mostafazadeh02-3}, so that all eigenvalues of our non-Hermitian Hamiltonian are real.
Therefore, our model never shows an exponential decay of the probability that one would find in dynamics of the Gorini-Kosakowski-Sudarshan-Lindblad  (GKSL) equation~\cite{GKS76,Lindblad76} under postselection of no quantum jumps~\cite{Ott16,Ashida16,Ashida17,Ashida20,Nakagawa20,Joglekar19,Joglekar21}.
We provide more details below.

Our model is not just a quantum toy model.
Maes \textit{et al.}~\cite{Maes22} showed in their paper how Dirac electrons can be analytically related to run-and-tumble particles by analyzing the equation of motion and numerically observing the interference pattern induced by a double slit both with Dirac electrons and run-and-tumble particles.
It is also known that the continuum limit of discrete-time quantum walks in one and two dimensions yields the Dirac equation~\cite{Strauch06,Yamagishi23}, as we show in Supplementary Materials S-I.
Thus, in a similar manner to the work by Maes \textit{et al.}~\cite{Maes22}, by unraveling our model, we may be able to relate our model to some classical active particle models.

\subsection*{Classical active Brownian particle}

Let us next review previous research on a prototypical classical active Brownian particle~\cite{Schweitzer98}.
This model may differ from major models used in the studies of active matter~\cite{Geier05, Howse07, Paxton04}.
However, since the authors of the study very clearly demonstrate the correspondence between the internal energy of the particle and its dynamics, we chose this study as the starting point of our discussion.
Schweitzer \textit{et al.}~\cite{Schweitzer98} studied the dynamics of a Brownian particle with the ability to take up energy from the environment, store it inside, convert internal energy into kinetic energy and move.
To model the dynamics, they added a new term of the internal energy $e(t)$ to the right-hand side of the Langevin equation.
They first studied the dynamics of an active Brownian particle under a harmonic potential with a constant energy take-up.
The active Brownian particle moves almost on a limit cycle with finite energy take-up, whereas a simple Brownian particle without energy take-up does not.
We aim to reproduce similar results: a quantum particle moves around more actively, climbing the harmonic potential with finite energy take-up.
Moreover, we aim to observe its quantum features that are absent in its classical counterpart.
To achieve these goals, we use nonunitary quantum walks~\cite{Mochizuki16,Xiao17,Hatano21b,Jiang24} as a tool.

\subsection*{Quantum walks}

The quantum walk is a quantum analog of the random walk.
Nonetheless, we note that the quantum walk exhibits deterministic quantum dynamics without any stochasticity. 
Instead of stochastic fluctuations of the classical random walker, the quantum walker moves under quantum interference at each site, which deterministically governs the dynamics of the walker’s wave function.
Its classical limit may be achieved only after introducing decoherence or other additional effects.

Quantum walk was originally introduced by Aharonov \textit{et al.}~\cite{Aharonov93}, who first referred to it as the ``quantum random walk''.
Meyer~\cite{Meyer96} built a systematic model for the quantum walk and revealed its  correspondence to the Feynman path integral~\cite{Feynman65} of the Dirac equation.
Beginning with Farhi and Gutmann~\cite{Farhi98}, quantum walks have been well studied in the context of quantum information~\cite{Ambainis12, Asaka21}.
To date, studies of quantum walks have become even more interdisciplinary and have extended to a variety of research fields, such as biophysics~\cite{Engel07, Dudhe22} and condensed-matter physics~\cite{Oka05}, particularly topological materials~\cite{Kitagawa10, Obuse11, Kitagawa12, Asboth13}.

The term "quantum walk" often refers to two types of time evolution, namely, continuous-time quantum walks and discrete-time quantum walks.
In the present paper, we focus on the latter, in which the space and time are both discrete.
In the next section, we investigate a nonunitary quantum walk as a quantum active particle in one dimension.
To extend our quantum active matter to higher dimensions, we utilize a quantum active particle in two-dimensional quantum walks~\cite{Yamagishi23}, using which we can obtain correspondence to the Dirac equation and further to the Schr\"{o}dinger equation in higher dimensions.
We present the results in Supplementary Materials S-V.
The final section summarizes the paper.
To make the paper self-contained, we provide a compact review of the quantum walk in the Supplementary Materials S-I.


\begin{figure}
  \centering
  \includegraphics[width=\textwidth]{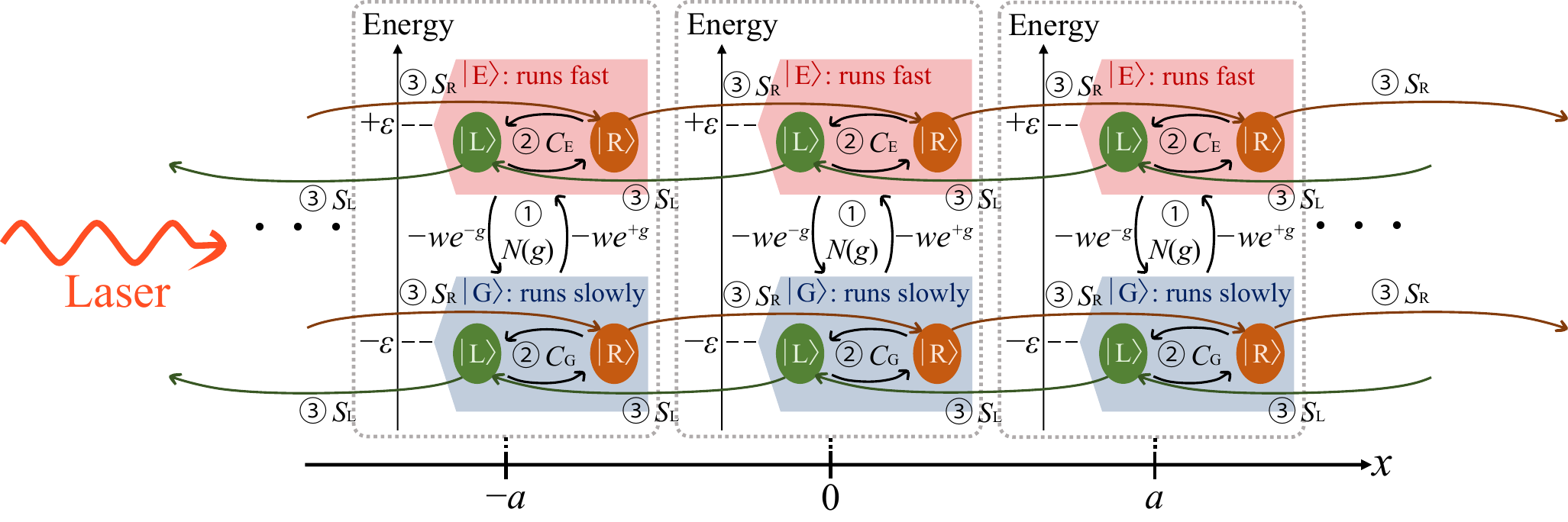}
  \caption{Time evolution of our quantum active particle in one dimension. There are four internal states, namely, $(\lk\oplus\rk)\otimes(\gs\oplus\es)$ at each site. The parameters $\varepsilon$, $w$ and $g$ are all real. We have the coin operator (\ctext{2}), the shift operator (\ctext{3}), and the new operator for the asymmetric transition between the ground state $\gs$ and the excited state $\es$, which describes the energy take-up (\ctext{1}). We use different $\theta$ values for the ground state $\gs$ and the excited state $\es$ to realize a system without momentum conservation.}
  \label{qabp_operators}
\end{figure}
\section*{Quantum active particle in one dimension}

\subsection*{One-dimensional model}

In one dimension, our quantum active particle has four internal states in total, namely, $(\lk\oplus\rk)\otimes(\gs\oplus\es)$.
Here, $\lk$ and $\rk$ denote the leftward and rightward states, respectively, whereas $\gs$ and $\es$ denote the ground and excited states, respectively.
We define the time evolution of our one-dimensional quantum active particle $\ket{\psi(T)}=[{U(g)}]^T\ket{\psi(0)}$ for $T\in\mathbb{Z}$ in terms of the following operators:
\begin{align}\label{eqQABP10}
N(g)&:=\sum_{x}\qty[
\dyad{x}{x}\otimes \ee^{-\ii H_{\NH}(g)}
],
\\\label{eqQABP11}
C&:=\sum_{x}\qty[
\dyad{x}{x}\otimes \ee^{-\ii H_C(x)}
],
\qquad
S:=\sum_{x}\qty[
\dyad{x-a}{x}\otimes \dyad{\mathrm{L}}
+\dyad{x+a}{x}\otimes \dyad{\mathrm{R}}
],
\end{align}
with $U(g):=SCN(g)$ and with the lattice constant $a$. Here,
\begin{align}\label{eqQABP10-1}
&H_{\NH}(g):=\qty(
\begin{array}{cc|cc}
-\varepsilon & 0 & -w \ee^{-g} & 0 \\
0 & -\varepsilon & 0 & -w \ee^{-g} \\
\hline
-w \ee^{+g} & 0  & +\varepsilon & 0 \\
0 & -w \ee^{+g} & 0 & +\varepsilon \\
\end{array}
),
\\\label{eqQABP11-1}
&H_C(x):=\qty(
\begin{array}{cc|cc}
0 & -\ii\theta_{\mathrm{G}}(x) & 0 & 0 \\
\ii\theta_{\mathrm{G}}(x) & 0 & 0 & 0 \\
\hline
0 & 0  & 0 & -\ii\theta_{\mathrm{E}}(x) \\
0 & 0 & \ii\theta_{\mathrm{E}}(x) & 0 \\
\end{array}
),
\quad
\dyad{\mathrm{L}}= 
\mqty(1 & 0 & 0 & 0 \\
0 & 0 & 0 & 0 \\
0 & 0 & 1 & 0 \\
0 & 0 & 0 & 0
),\quad
\dyad{\mathrm{R}}=
\mqty(0 & 0 & 0 & 0 \\
0 & 1 & 0 & 0 \\
0 & 0 & 0 & 0 \\
0 & 0 & 0 & 1
).
\end{align}
We used the bases $\{\lk\otimes\gs, \rk\otimes\gs, \lk\otimes\es, \mathrm{and} \rk\otimes\es\}$ in this order to represent the matrices.
In equation~\eqref{eqQABP11}, the shift operator $S$ is equivalent to $\exp(-\sigma_z \dd/\dd x)$ and hence provides a kinetic-energy contribution for the Hamiltonian.
In equation~\eqref{eqQABP10-1}, $\mp\varepsilon$ denotes the levels of $\ket{\mathrm{G}}$ and $\ket{\mathrm{E}}$, respectively, whereas the non-Hermitian parameter $g$ specifies the difference in the transitions between the two levels.
The non-Hermiticity of $H_\NH$ makes the total time-evolution operator $U(g)$ nonunitary.
We can interpret the non-Hermiticity of $H_\NH$ as an effect of laser pumping, as discussed in the next subsection. 
Note that the energy conservation is broken because of our non-Hermitian Hamiltonian $H_\NH(g)$. 
Thus, it satisfies the property (i) of quantum active matter; namely, the energy is not conserved.
In fact, we will show below in equation~\eqref{eqQABP17} that the Hamiltonian~\eqref{eqQABP10-1} has a symmetry called pseudo-Hermiticity~\cite{Mostafazadeh02-1,Mostafazadeh02-2,Mostafazadeh02-3}; hence, the energy eigenvalues remain real, not depending on $g$ at all, but the energy expectation value is not conserved because the eigenvectors are not orthogonal to each other.

In equation~\eqref{eqQABP11-1}, we set the parameter for the excited state $\theta_{\mathrm{E}}(x)$ to be generally less than that of the ground state $\theta_{\mathrm{G}}(x)$.
This is based on the following.
The continuum limit of the unitary time evolution of one-dimensional~\cite{Strauch06} and two-dimensional~\cite{Yamagishi23} quantum walks yields a Dirac Hamiltonian with the parameters for the coin operator $\theta$ of the former being proportional to the mass terms of the latter.
We set $\theta_{\mathrm{E}}<\theta_{\mathrm{G}}$ so that our quantum active particle moves faster in the excited state than in the ground state.
In other words, our active quantum walker does not conserve the momentum, which is the property (ii) of quantum active matter; the momentum is also not conserved.
Since the particle moves faster when pumped from the ground state to the excited state, it also satisfies the property (iii) of quantum active matter; the kinetic motion depends on the internal state of the particle.
See Fig.~\ref{qabp_operators} for details of the time evolution.

It is important to note that we can transform the non-Hermitian matrix $H_{\NH}(g)$ into a Hermitian matrix using a similarity transformation called the imaginary gauge transformation~\cite{Hatano96, Hatano97}
$A(g)=\mathop{\mathrm{diag}}(\ee^{-g/2}, \ee^{-g/2}, \ee^{+g/2}, \ee^{+g/2})$,
as in
\begin{align}\label{eqQABP17}
A(g)^{-1}H_{\NH}(g)A(g)%
=A(-g)H_{\NH}(g)A(g)%
= \qty(
\begin{array}{cc|cc}
-\varepsilon & 0 & -w & 0 \\
0 & -\varepsilon & 0 & -w \\
\hline
-w & 0 & +\varepsilon & 0 \\
0 & -w & 0 & +\varepsilon \\
\end{array}
)%
= H_{\NH}(g=0).
\end{align}
Note also that both $S$ and $C$ commute with $A(g)$;
the matrix $A(\pm g)$ consists of two blocks for the ground and excited states, each of which is just the identity matrix multiplied by either $\ee^{\pm g/2}$, whereas the matrices $S$ and $C$ also consist of two blocks.
Therefore, the total time-evolution operator $U(g)$ can also be transformed to $U(0)$, as in
$A(g)^{-1}U(g)A(g)=U(0)$;
therefore, the eigenvalues of the nonunitary matrix $U(g)$ are identical to those of the unitary matrix $U(g=0)$, which are located on the unit circle in the complex plane.
In other words, the eigenvalues of our Hamiltonian~\eqref{eqQABP10-1} remain real for any values of the non-Hermiticity parameter $g$.
This property is often called pseudo-Hermiticity~\cite{Mostafazadeh02-1,Mostafazadeh02-2,Mostafazadeh02-3}.
We may call the corresponding property of $U(g)$ pseudounitarity.
Because there is no energy dissipation, we can refer to our dynamics as purely deterministic quantum dynamics;
see subsection~``Two features of the pseudo-Hermiticity: all real eigenvalues and nonorthogonal eigenvectors" below.

\subsection*{Non-Hermiticity and laser pumping}

We can understand the physical meaning of the parameter $g$ in the non-Hermitian Hamiltonian in equation~\eqref{eqQABP10-1} 
in terms of the rate equations for a two-level system, as discussed in Refs.~\cite{LightI81,LightII85}.

Let us consider $N_1+N_2$ pieces of two-level systems under external laser pumping with the occupation number of atoms $N_1$ and $N_2$ in the ground and excited states, respectively; the total number of atoms is fixed to $N=N_1+N_2$.
In the case with no stimulated emission, we obtain the rate equations for the occupation number of the two states as follows:
\begin{align}
&\dv{N_1}{t}=w_{12}N_2-w_{21}N_1+(N_2-N_1)Wn, \label{eqQABP30} \\
&\dv{N_2}{t}=w_{21}N_1-w_{12}N_2-(N_2-N_1)Wn. \label{eqQABP31}
\end{align}
Here, $w_{21}$ is the transition rate of the photons from the ground state to the excited state due to external pumping, and $w_{12}$ is the decay rate from the excited state to the ground state due to spontaneous emission.
We let $n$ denote the number of photons in the environment and $W$ denote the transition rate of decay to the environment.
We can determine the correspondence between the two rates and the elements of the Hamiltonian $H_\NH$ in equation~\eqref{eqQABP10-1}, 
via Fermi's golden rule as follows~\cite{QuantOpt92}:
$w_{21}=\abs{\mel{\mathrm{E}}{H_{\NH}}{\mathrm{G}}}^2=w^2\ee^{2g}$, 
$w_{12}=\abs{\mel{\mathrm{G}}{H_{\NH}}{\mathrm{E}}}^2=w^2\ee^{-2g}$. 

Subtracting equation~\eqref{eqQABP31} from equation~\eqref{eqQABP30} yields
\begin{align}\label{eqQABP34}
\dv{D}{t}=(w_{21}-w_{12})N-(w_{21}+w_{12})D-2WnD
\end{align}
with
$N:=N_1+N_2=\mbox{const.}$ and 
$D:=N_2-N_1$. 
In the stationary state for which the left-hand side of the equation above vanishes, we have
\begin{align}
D=\frac{w_{21}-w_{12}}{w_{21}+w_{12}+2Wn_0}N \label{eqQABP37}
\end{align}
with a fixed number of photons in the environment $n_0$; the whole system is pumped by a laser, and $n$ stays constant.
From the conditions for $N$ and $D$ expressed in equation~\eqref{eqQABP37}, we obtain the ratio between the occupation numbers of the two states in the stationary state as
\begin{align}\label{eq35}
\frac{N_2^\st}{N_1^\st}%
=\frac{N+D}{N-D}=\frac{w_{21}+Wn_0}{w_{12}+Wn_0}
=\frac{w^2\ee^{2g}+Wn_0}{w^2\ee^{-2g}+Wn_0}
\simeq\frac{w^2\ee^{2g}}{Wn_0}\propto\ee^{2g}
\end{align}
for the stationary values $N_1^\st$ and $N_2^\st$ for $N_1$ and $N_2$, respectively.
We used the relation $w^2\ee^{-2g}\ll Wn_0\ll w^2\ee^{2g}$ with $g$ assumed to be finite.
We can thus assume that the parameter $g$ describes the probability difference between the two levels of a qubit under laser pumping.
Since $g$ does not depend on $\varepsilon$ in equation~\eqref{eqQABP10}
, these two parameters are set independently of each other.

Because of the energy input of the laser pumping, the total probability and the energy expectation of our model should change over time. 
This physical requirement is satisfied by the pseudo-Hermiticity of our model.
We explain this point in the next subsection.

\subsection*{Two features of the pseudo-Hermiticity: all real eigenvalues and nonorthogonal eigenvectors}

At the end of the subsection~``One-dimensional model", we stressed that our model does not include any explicit dissipation.
In this sense, our non-Hermitian Hamiltonian~\eqref{eqQABP10-1} is distinct from non-Hermitian Hamiltonians that one would usually find in the GKSL equation~\cite{GKS76,Lindblad76} under postselection of no quantum jumps~\cite{Ott16,Ashida16,Ashida17,Ashida20,Nakagawa20,Joglekar19,Joglekar21}.

The GKSL equation for the density operator $\rho_0$ of an open quantum system can be written in the following form:
\begin{align}
\ii\dv{t}\rho_0 =  \qty(H_\textrm{GKSL}\rho_0-\rho_0 H_\textrm{GKSL}^\dag)+\ii \gamma L\rho_0 L^\dag
\end{align}
with the non-Hermitian Hamiltonian
\begin{align}
H_\textrm{GKSL}=H_0-\ii\frac{\gamma}{2}L^\dag L,
\end{align}
where $H_0$ is a Hermitian Hamiltonian of the system of interest, $L$ is a jump operator representing the effects of the environment, and $\gamma$ is a positive constant denoting dissipation to the environment.
Stochastic unraveling~\cite{Daley14,Landi24} of the GKSL equation describes its dynamics as follows.
The non-Hermitian Hamiltonian $H_\textrm{GKSL}$ enforces exponential decays because of the negative imaginary parts of its complex eigenvalues due to the factor $-\ii L^\dag L$.
A quantum jump due to the term $\ii \gamma L\rho L^\dag$ randomly kicks in and revives the system out of the decay.
Then, the exponential decay resumes immediately after the jump.
Each set of stochastic quantum jumps defines a quantum trajectory.
Averaging the quantum trajectories over all possible occurrences of quantum jumps, we find that the total probability $\mathop{\textrm{Tr}}\rho$ is conserved.

By contrast, all eigenvalues of our Hamiltonian given in equations~\eqref{eqQABP10-1}--\eqref{eqQABP11-1} are real because of the pseudo-Hermiticity; hence,
there will be no exponential decay if we considered stochastic unraveling of our dynamics.
On the other hand, the total probability fluctuates in time because of another feature of the pseudo-Hermiticity, namely the lack of orthogonality among the eigenvectors, as we explain below.

We define the total probability as 
\begin{align}
\label{eqQABP18}
P_\textrm{tot}(T)=\sum_xP(x,T)=\braket{\psiR(T)}{\psiR(T)},
\qquad\mbox{where}\quad
P(x,T)&:=\abs{\braket{x}{\psiR(T)}}^2.
\end{align}
Note that we use the right-eigenvector $\ket{\psiR}$ of our non-Hermitian Hamiltonian and its Hermitian conjugate $\bra{\psiR}:=\ket{\psiR}^\dag$ to calculate the probability density, as would be done in standard quantum mechanics. 
If we regarded our system as a closed non-Hermitian system, we would use the left-eigenvector $\bra{\psiL}$ instead of $\bra{\psiR}$~\cite{Brody16}.
Since we regard our system as an open quantum system, we use the standard definition of the expectation value; see Appendix A of Ref.~\cite{Hatano24}.
From equation~\eqref{eqQABP17}, we can obtain that a right-eigenvector of the pseudo-Hermitian Hamiltonian $H_\textrm{NH}(g)$ is given by the corresponding right-eigenvector of the Hermitian Hamiltonian $H_\textrm{NH}(g=0)$ multiplied by $A(g)$, whereas a left-eigenvector of the former is given by the left-eigenvector, which is the Hermitian conjugate of the right-eigenvector in the Hermitian case multiplied by $A(g)^{-1}$. 
Therefore, in general, there is no orthogonality among the right-eigenvectors for $g\neq0$: $\braket{\psiR_m}{\psiR_n}\neq\delta_{mn}$;
there is only a biorthogonality: $\braket{\psiL_m}{\psiR_n}=\delta_{mn}$.
Nonetheless, we use $\bra{\psiR}$ and $\ket{\psiR}$ because we regard our model as an open quantum system; see again Appendix A of Ref.~\cite{Hatano24}.

Defining the probability and the energy expectation value in this manner results in their fluctuations in time.
Let us expand an initial state in terms of the right-eigenvectors as $\ket{\psi(0)}=\sum_n c_n \ket{\psiR_n}$.
The time evolution of the state is given by $\ket{\psi(T)}=\sum_n c_n \ee^{-\ii E_n T}\ket{\psiR_n}$.
The total probability given in equation~\eqref{eqQABP18} is then expressed by
\begin{align}
P_\textrm{tot}(T)=\braket{\psiR(T)}{\psiR(T)}=
\sum_{m,n}c_m^\ast c_n \ \ee^{-\ii(E_n-E_m)T}\braket{\psiR_m}{\psiR_n}.
\end{align}
In the standard quantum mechanics, the right-hand side is reduced to a constant $\sum_n \abs{c_n}^2$ because of the orthonormality of the eigenvectors, so that the total probability is conserved. In the present case of the pseudo-Hermiticity, a lack of orthogonality of the eigenvectors leads to the fluctuations of the total probability with time.
The same logic can also demonstrate the fluctuation of the energy expectation values;
see Supplementary Material S-II for details. 
See also Fig.~S4 for the actual fluctuation in the total probability.

Summarizing the discussions in the previous and present subsections,  two features of the pseudo-Hermiticity make the dynamics of our model distinct from the dynamics due to the GKSL equation.
First, the reality of all eigenvalues makes the dynamics devoid of explicit gain or loss that the GKSL equation has.
Nonetheless, the lack of orthogonality among the eigenvectors makes the total probability and the energy expectation values fluctuate over time.
We will see numerically below that their temporal fluctuations neither converge nor diverge exponentially. 

\subsection*{Numerical results for the one-dimensional model}

Here, we present our results of numerical calculation results in one dimension.
Starting with computation on a flat line, we first examine the basic properties of our model.
We present the results for the topological edge states 
and for an effective harmonic potential 
later in the present section.
We set $\hbar=a=1$ for all numerical calculations.

We define the normalized probability distribution as
\begin{align}
\label{eqQABP18-2}
\tilde{P}(x, T)%
&:=\frac{P(x,T)}{\sum_x P(x,T)}.
\end{align}
We normalize the probability here because $\sum_x P(x,T)$ is not conserved for a finite value of $g$.
We use the normalized probability~\eqref{eqQABP18-2} hereafter.
We also define the mean position $\expval{x(T)}$ and the standard deviation $\Delta x(T)$ of the walker at each time step as
\begin{align}
\expval{x(T)}:=\sum_x{x \tilde{P}(x, T)},\quad
\Delta x(T):=\sqrt{\sum_x\mqty[(x-\expval{x(T)})^2 \tilde{P}(x, T)]}, \label{eqQABP16}
\end{align}
respectively.

\subsubsection*{Dynamics on a flat potential}

We first examine the dynamics of our quantum active particle without any potentials, namely, on a flat line.
Let us set the parameters for the coin operator \eqref{eqQABP11}, $\theta_{\mathrm{G}}$ and $\theta_{\mathrm{E}}$, to the following constants:
\begin{align}\label{eqdelta1}
\theta_{\mathrm{G/E}}(x)=
\theta_{\mathrm{g/e}}:=\theta_0\pm\varepsilon.
\end{align}
As explained after equation~\eqref{eqQABP11-1}, we choose this parameter so that $\theta_{\mathrm{e}}<\theta_{\mathrm{g}}$.
Unless noted otherwise, all calculations 
for the flat potential were conducted with the parameter values fixed as follows:
\begin{align}\label{eqdelta2}
\theta_0=\frac{\pi}{3},\quad\varepsilon=0.25,\quad w=0.25,
\end{align}
and the system size is $L_x=401$ with $-200\le x\le200$.

\begin{figure}
  \centering
  \includegraphics[width=\columnwidth]{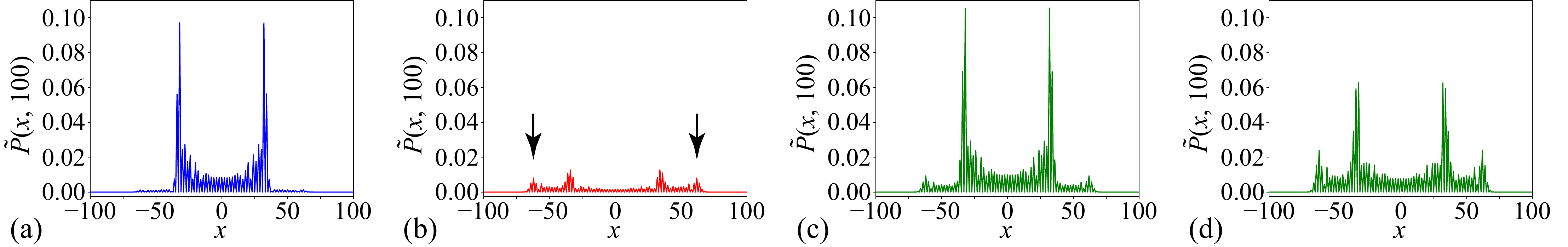}
  \caption{Normalized probability distributions $\tilde{P}(x,T=100)$ of the ground state [(a)], excited state [(b)] and the sum of the both states [(c), (d)] after 100 time steps of evolution for $g=0$ [(a), (b), (c)] and $g=1$ [(d)]. The system size is $L_x=401$ with $-200\le x\le200$; the probability outside the plotting range is significantly small.}
  \label{qabp_28delta_ani}
\end{figure}

Figure~\ref{qabp_28delta_ani} shows the normalized probability distributions $\tilde{P}(x,T)$  after 100 time steps of evolution from the initial condition of the delta peak only in the ground state at the origin site.
The time evolution of the ground state shown in Fig.~\ref{qabp_28delta_ani}(a) is a typical probability distribution of quantum walks.
In quantum walks, all possible paths leading to the origin site should interfere with each other and be cancelled out, which yields the minimum around the origin.
On the other hand, the paths leading to the wave fronts on the left and the right sides should be fewer in number and be cancelled less, which yields the maximum peaks around the wave fronts.

On the other hand, the normalized probability distribution of the time evolution in the excited state shown in Fig.~\ref{qabp_28delta_ani}(b) has a more complicated structure.
The peaks on the outside (indicated by the arrows) arise because our quantum-active particle moves faster in the excited state than in the ground state.
The peaks on the inside are due to the transition from the peaks of the ground state, in which the particle moves more slowly than in the excited state.
When we further turn on the non-Hermitian activity parameter $g$, the probability for the excited state is increased from that in Fig.~\ref{qabp_28delta_ani} (c) to the higher value in Fig.~\ref{qabp_28delta_ani} (d).
This clearly demonstrates the dynamics that we illustrate in Fig.~\ref{qabp_operators}.

Let us quantitatively discuss the speed of the motion of our quantum walker.
In Fig.~\ref{qabp_28delta_ani} (a) and (b), the peak locations for the ground state are $\pm 32$ after 100 time steps, and those for the excited state are $\pm 62$. 
In the absence of coupling between the ground and excited states, the front peaks of each state move at the group velocity $\cos \theta_{\mathrm{g/e}}$~\cite{Kempf09}; see Supplementary Materials S-III.
In the particular case of the parameter values in equation~\eqref{eqdelta1} with equation~\eqref{eqdelta2}, we would have
$\cos\theta_\mathrm{g}\simeq 0.270$ and
$\cos\theta_\mathrm{e}\simeq 0.698$.
When we introduce a coupling between the ground and excited states, however, we numerically find for the parameter values $\varepsilon=w=0.25$ (see also Supplementary Materials S-III) that the maximum group velocity for the ground state is approximately $0.343$ at $k\simeq 0.624\pi$ and that for the excited state it is approximately $0.642$ at $k\simeq0.450\pi$.
This finding is consistent with the numerical observations ($0.32$ and $0.62$, respectively) in Figs.~\ref{qabp_28delta_ani} (a) and (b).
This result indicates that the faster running peaks of the excited state are dragged down by the more slowly running peaks of the ground state, whereas the latter are pulled up by the former.

\begin{figure}
  \centering
  \includegraphics[width=\columnwidth]{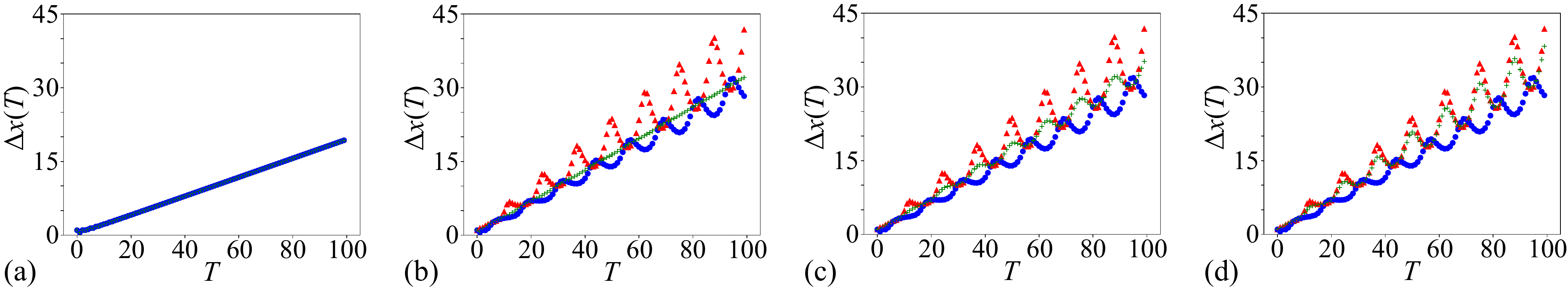}
  \caption{The time-step dependence of the standard deviation $\Delta x$ for $w=0$ [(a)] and $w=0.25$ [(b), (c), (d)]. (a) and (b) are computed for $g=0$, (c) is computed for $g=0.5$, and (d) is computed for $g=1$. The red triangles and blue circles indicate the standard deviation with respect to the excited and ground states, respectively. The green plus symbols indicate the standard deviation normalized by the total probability.}
  \label{qabp_28delta_SD}
\end{figure}

As discussed above regarding non-Hermiticity and laser pumping,
our quantum active particle, which has two levels, is pumped from the lower level to a higher level by an external laser.
Hence, there exists an oscillation between the ground and excited states.
Figure~\ref{qabp_28delta_SD} shows the time dependence of the standard deviation calculated for different parameter values.
First, Fig.~\ref{qabp_28delta_SD} (a) shows the case in which there is no coupling between the ground and excited states: $w=0$.
In this case, the standard deviations of both the ground and excited states as well as their sum converge to one curve; since there is no transition between the two states, we do not observe an oscillation.

When $w$ is set to finite values, as shown in Figs.~\ref{qabp_28delta_SD} (b)--(d), the standard deviations of both the ground and excited states oscillate with a relative phase shift $\pi$, which implies a resonant transition between two states.
This resonant transition is one of the quantum features that we observe in our quantum active matter.
For a detailed discussion of the time period of the oscillations $T_{\mathrm{osc}}$; see Supplementary Materials S-IV.\\[-\baselineskip]
\begin{wrapfigure}[29]{r}{0.494\columnwidth}
\vspace{-1.7\baselineskip}
  \centering
    \includegraphics[width=0.494\columnwidth]{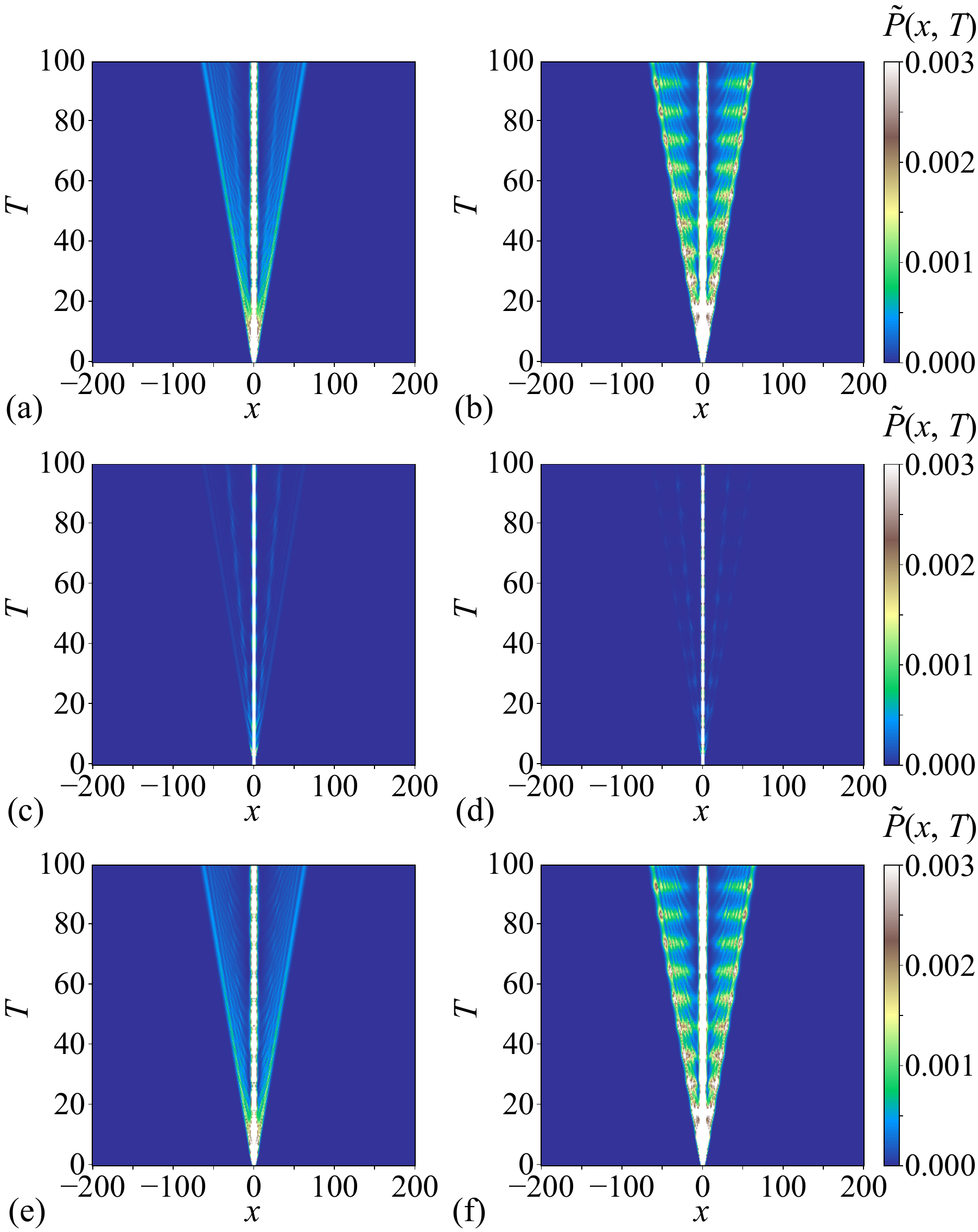}
    \caption{Density plot of the time evolution of the quantum walker focusing on the ground state [(c), (d)], the excited state [(e), (f)] and the sum of the two states [(a), (b)] for $g=0$ [(a), (c), (e)] and $g=1$ [(b), (d), (f)]. The color indicates the probability of the walker at each site and time step.}
    \label{qabp_28step_Dens}
\end{wrapfigure}
We note that the standard deviation normalized by the total probability, which is plotted with green plus symbols, does not oscillate in the case of $g=0$ (Fig.~\ref{qabp_28delta_SD} (b)), but does oscillate for  $g>0$ (Fig.~\ref{qabp_28delta_SD} (c), (d)).
This is because as we increase $g$, the ratio of the excited state increases as $\ee^{2g}$ and physical quantities such as the means and the standard deviations of the total system tend to take closer values to those of the excited state.
The oscillations observed in Fig.~\ref{qabp_28_Mean_SD} (c), (d) and Fig.~S7 (b) in the Supplementary Materials S-V can be explained in the same manner.

\subsubsection*{Topological edge states}

Next, we study the edge states of our quantum active particle.
Here, we set the parameters for the coin operator \eqref{eqQABP11}, $\theta_{\mathrm{G}}$ and $\theta_{\mathrm{E}}$, as
\begin{align}\label{eqedge1}
\theta_{\mathrm{G/E}}(x)=\begin{cases}
\theta_{\mathrm{g/e}} & \mbox{for $x\le0$} \\
-\theta_{\mathrm{g/e}} & \mbox{for $x>0$}
\end{cases},
\quad\theta_{\mathrm{g/e}}:=\theta_0\pm\varepsilon.
\end{align}
Therefore, edge states may emerge at the discontinuities of the parameter values at the origin $x=0$ and at the edges of the system $x=\pm L/2$, where $L$ denotes the system size, and we assume periodic boundary conditions.
The appearance of these states bound to the edges was predicted in Refs.~\cite{Jackiw76, Yamagishi23} for the Dirac Hamiltonian in one dimension; note that the quantum walk converges to the Dirac system in the continuum limit~\cite{Strauch06, Yamagishi23}.
Since the potential of equation~\eqref{eqedge1} squared is constant, the states other than the edge states propagate in the same manner as those 
for the flat potential.

We first set the initial state to the edge state at $x=0$ of the ground state under the condition of no transition between the ground and excited states.
We then let the state evolve in time as the transition $w$ is turned on.
We thereby expect that the edge state of the ground state may remain bounded to the edge, whereas the component pumped up to the excited state may escape away from the edge.
The time evolution in Fig.~\ref{qabp_28step_Dens} precisely reproduces our expectation.
All of the computations of 
the topological edge states were conducted with the parameter values fixed as follows:
\begin{align}\label{eqedge2}
\theta_0=\frac{\pi}{3},\quad\varepsilon=0.25,\quad w=0.25.
\end{align}
In Fig.~\ref{qabp_28step_Dens}(c), most of the wave functions of the ground state remain around the origin $x=0$, whereas some of the components pumped up to the excited state move away from the edge, as shown in Fig.~\ref{qabp_28step_Dens}(e).
Note that a small fraction of the ground state that leaks away in Fig.~\ref{qabp_28step_Dens}(c) is in fact the component that transitions back down from the excited-state component escaping to the outside and then runs behind the excited state.

As we turn on the activity parameter $g$, a significant fraction of the excited state escapes away from the bound of the edge, as shown in Fig.~\ref{qabp_28step_Dens}(f).
This indeed demonstrates that the quantum walker is activated by the parameter $g$.

\begin{wrapfigure}[16]{r}{0.463\columnwidth}
  \centering
  \includegraphics[width=0.463\columnwidth]{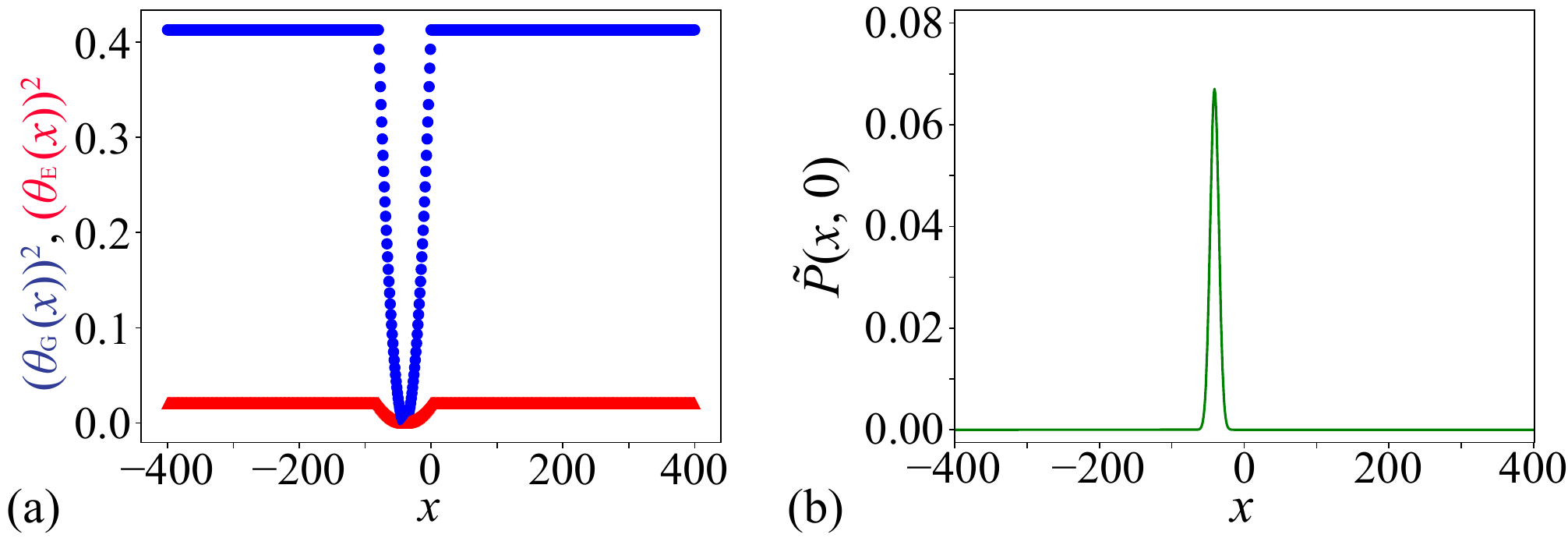}
  \caption{(a) Potential of equation~\eqref{eqQABP13} squared with the parameters in equation~\eqref{eqQABP15} and $L_x=801$. The vertical axis indicates the squared potentials for the ground state (blue markers) and the excited state (red markers). The horizontal axis indicates the space.
  (b) An eigenstate of the eigenvalue close to unity, $U_n=0.9378050525983931+\ii\,0.3471623299278579$, with the parameters in equation~\eqref{eqQABP15} and $L_x=801$ and $g=0$, which we found numerically. The eigenstate is shifted $\delta_x=19$ steps to the right of the initial state for the computation of dynamics.}
  \label{qabp_28_L=801}
\end{wrapfigure}
%
\subsubsection*{One-dimensional oscillator}

We finally investigate the dynamics of our quantum active particle under an effective harmonic potential.
For this purpose, we set 
the parameters for the coin operator \eqref{eqQABP11}, $\theta_{\mathrm{G}}$ and $\theta_{\mathrm{E}}$, linear in $x$:
\begin{align}\label{eqQABP13}
&\theta_{\mathrm{G/E}}(x)=\begin{cases}
\theta_{\mathrm{g/e}} & \mbox{for $x<\frac{-\alpha-1}{\beta}$} \\
\theta_{\mathrm{g/e}}(\alpha+\beta x) & \mbox{for $\frac{-\alpha-1}{\beta}\le x\le\frac{-\alpha+1}{\beta}$} \\
-\theta_{\mathrm{g/e}} & \mbox{for $x>\frac{-\alpha+1}{\beta}$}
\end{cases},\notag \\
&\theta_{\mathrm{g/e}}:=\theta_0\pm\varepsilon.
\end{align}
While the Dirac particle in the continuum limit of the quantum walk~\cite{Strauch06} perceives the linear potential in equation~\eqref{eqQABP13},  the corresponding Schr\"{o}dinger particle perceives~\cite{Yamagishi23} the potential squared, which is a harmonic potential in the region $(-\alpha-1)/\beta\le x\le(-\alpha+1)/\beta$.
All computations 
for the one-dimensional oscillator were conducted with the parameters fixed as follows:
\begin{align}\label{eqQABP15}
\theta_0=\frac{\pi}{8},\quad\varepsilon=w=0.25,\quad\alpha=1,\quad\beta=0.025.
\end{align}

\begin{wrapfigure}[21]{r}{0.28\columnwidth}
\vspace{-\baselineskip}
  \centering
    \includegraphics[width=0.28\columnwidth]{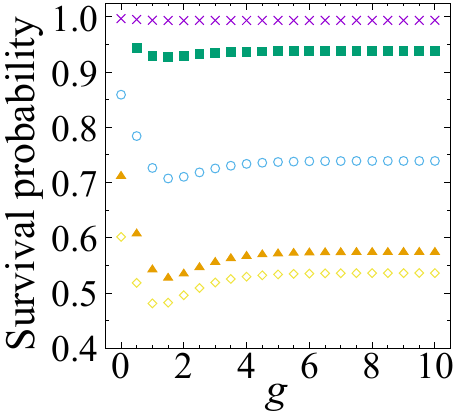}
    \caption{Dependence of the survival probability in the potential in the region of $-80\le x\le 0$, on $\delta_x$ and $g$. We plot the survival probability for $\delta_x=9, 14, 19, 24\ \mbox{and}\ 29$ with purple crosses, green-filled squares, blue circles, orange-filled triangles and yellow diamonds, respectively.}
    \label{qabp_28_survivalprob}
\end{wrapfigure}
\noindent
Figure~\ref{qabp_28_L=801}(a) shows the potential of equation~\eqref{eqQABP13} squared,
which is perceived by the corresponding Schr\"{o}dinger particle in the continuum limit (cf.~Ref.~\cite{Yamagishi23}) perceives.
For the initial state for the computation, we shift to the right by $\delta_x$ sites an eigenstate of the eigenvalue close to unity with the parameters in equation~\eqref{eqQABP15} and $g=0$, which we found numerically; see Fig.~\ref{qabp_28_L=801}(b).\

We first investigate the dependence of the survival probability, that is, the probability of being trapped by
the potential that spans the region of $-80\le x\le 0$, on $\delta_x$ and $g$; see Fig.~\ref{qabp_28_survivalprob}.
(Here, the system size is $L_x=401$ with $-200\le x\le200$ under the periodic boundary conditions.)
Starting from $g=0$, we see that the survival probability decreases as $g$ increases to some point, but at approximately $g=2$ (depending on $\delta_x$, which is the number of steps by which we shift the eigenstate), it starts to increase slightly and converges to a constant up to $g=10$, for which the probability originates almost entirely from the excited state.

\begin{figure}
  \centering
  \includegraphics[width=\columnwidth]{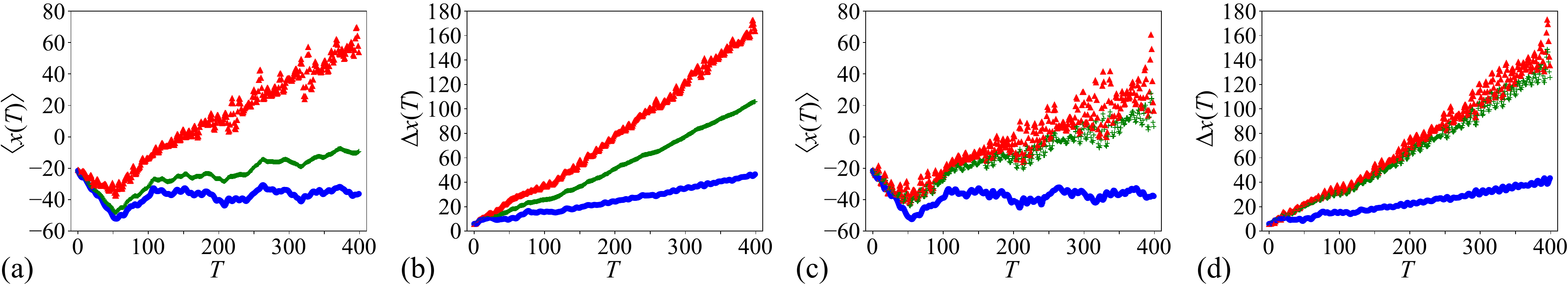}
  \caption{The mean position  [(a), (c)]  and the standard deviation [(b), (d)] of the movement of the quantum walker for $g=0$ [(a), (b)] and $g=1$ [(c), (d)]. The vertical axis indicates the mean position and the standard deviation, while the horizontal axis indicates the time steps. The red triangles and blue circles indicate the mean and the standard deviation with respect to the excited and ground states, respectively. The green plus symbols indicate the mean position and the standard deviation normalized by the total probability.}
  \label{qabp_28_Mean_SD}
\end{figure}

We then examine the dynamics of our quantum active particle more closely for a larger system size, $L_x=801$, with $-400\le x\le400$ under periodic boundary conditions.
For the initial state for the computation hereafter, we fix the shift to $\delta_x=19$ sites.
The mean position $\expval{x(T)}$ and the standard deviation $\Delta x(T)$ of the walker at each time step
are shown in Fig.~\ref{qabp_28_Mean_SD}.
The fluctuations in Fig.~\ref{qabp_28_Mean_SD}(c), (d) for $g=1$ arise from the nonconservation of probability.
Figures~\ref{qabp_28_Mean_SD}(b), (d) show that the standard deviation tends to take larger values as $g$ increases.
The difference for different values of $g$ is not only due to the change in the ratio between the ground and excited states; rather, in Fig.~\ref{qabp_28_Mean_SD}, we observe changes in each of the two states.

\begin{wrapfigure}[25]{r}{0.55\columnwidth}
\vspace{-\baselineskip}
  \centering
  \includegraphics[width=0.55\columnwidth]{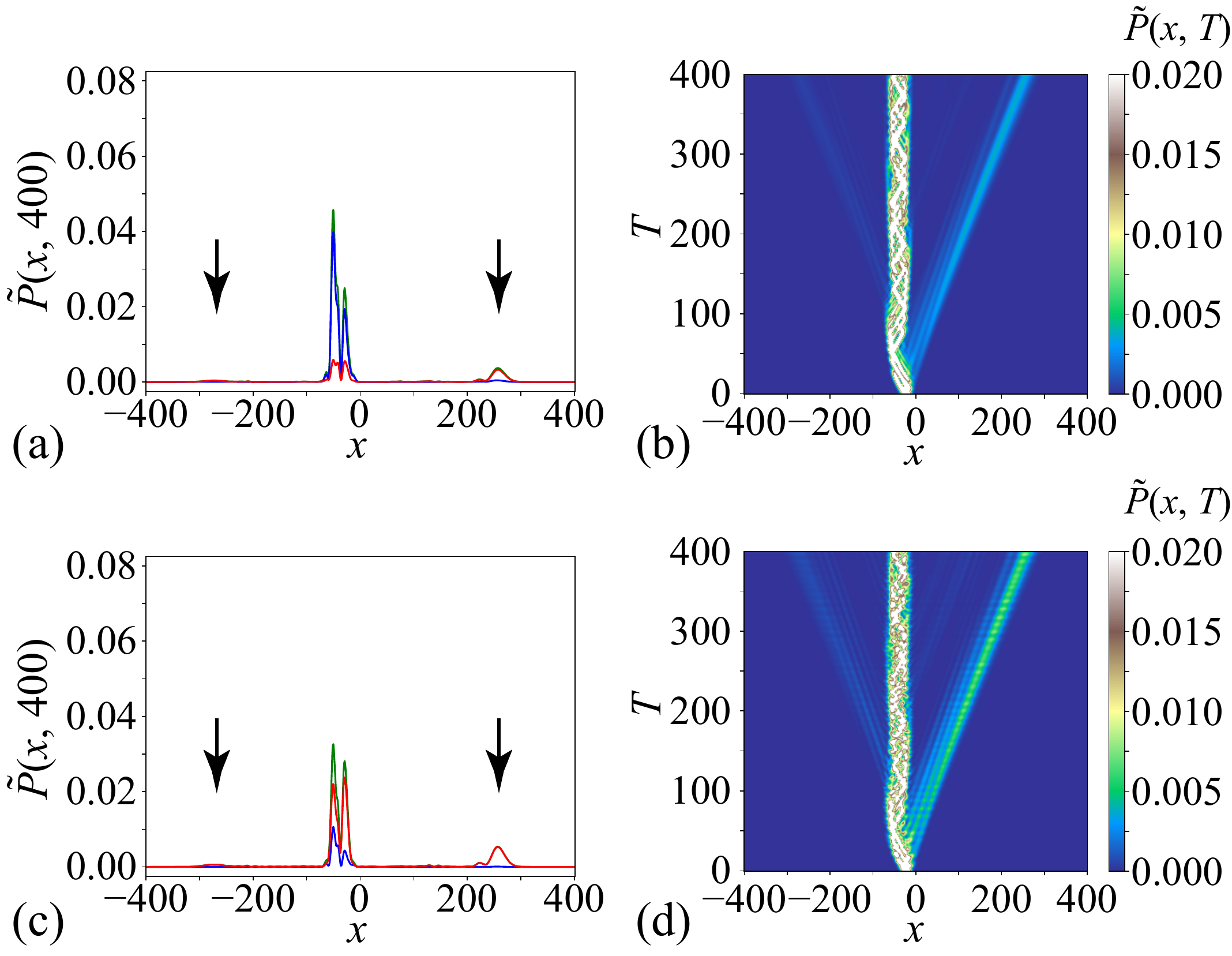}
  \caption{The normalized probability distribution $\tilde{P}(x,T=400)$. [(a), (c)] and the density plot of the time evolution [(b), (d)] of the quantum walker for $g=0$ [(a), (b)] and $g=1$ [(c), (d)]. For (a) and (c), the red curves and blue curves correspond to the excited and ground states, respectively, whereas the green curves indicate the sum of the probabilities for both states. For (b) and (d), the vertical axis indicates the time steps, and the horizontal axis indicates the site. The color indicates the probability of the walker at each site and time step.}
  \label{qabp_28_ani_400_Dens}
  \end{wrapfigure}

The normalized probability distributions of the ground and the excited states at each site after 400 time steps of evolution are shown in the left column of Fig.~\ref{qabp_28_ani_400_Dens}.
The side peaks indicated by the arrows become larger than the center peak as $g$ increases.
Based on this, we claim that we have succeeded in defining a quantum active particle using nonunitary quantum walks, namely we obtained results similar to those obtained in the previous research~\cite{Schweitzer98} on a classical active Brownian particle; that is, the particle becomes more active and increases in potential.
Note that our non-Hermiticity parameter $g$ corresponds to the energy-take-up term $q(\vb{r})$ in the previous research~\cite{Schweitzer98}.

Moreover, we also observed unique quantum features.
The right column of Fig.~\ref{qabp_28_ani_400_Dens} shows density plots of the time evolution.
We can clearly see the ballistic spreading of the side peaks of the excited state, particularly for $g=1$.
We see that some curves first emerge in the $x>0$ region.
This is because we shift the eigenstate of the time-evolution operator in the $x$ plus direction for the initial state.
Our quantum active particle hits the potential wall on the right side, which is located at $-40\le x\le0$ first, and then some portion climbs up the wall and escapes outside around $T\simeq10$.
Some portion is reflected on the right potential wall and then hits the potential on the left side, which is located at $-80\le x\le-40$.
Some portion then climbs up the potential wall and escapes outside around $T\simeq100$, and some portion is reflected.
The particle repeats the same behavior; more peaks are also separated from the central peak at approximately $T=200$ and $T=300$.

%
We also conducted numerical simulations in two dimensions, making the discrete-time quantum walk in two dimensions in Ref.~\cite{Yamagishi23} nonunitary.
We again observe that our quantum active particle moves around in a wider region with larger values of $g$.
This is a result similar to that of the previous study~\cite{Schweitzer98} of a classical system.
Moreover, we observe that the movement of our quantum active particle almost converges to a limit cycle for $g=0$, and the particle does not fall into the origin.
This is a quantum feature: our quantum active particle stays on a constant energy surface without energy take-up.
As another quantum feature, we again observe resonant two-state oscillation between the ground and excited states in standard deviations of $x$ and $y$.
See Supplementary Materials S-V for more details of the model and numerical results in two dimensions.

\section*{Summary}

In the present paper, motivated by a study~\cite{Schweitzer98} on an active Brownian particle, we have introduced a model of quantum active matter by making the quantum walk in one and two dimensions nonunitary.
Our quantum active matter allows us to study the real-time dynamics of the system in a fully quantum range without external manipulation~\cite{Zheng23}.
This type of quantum active matter was realized in our model for the first time.
Considering the three properties (i)--(iii) of quantum active matter listed in the Introduction,
we expect that our research will open up new research directions in two ways, as shown in Table~\ref{tab1}.

We introduced a new internal degree of freedom, namely, the energy ground state $\gs$ and excited state $\es$, to realize a quantum system without energy or momentum conservation.
Together with these new internal states, we also introduce a new nonunitary operator $N(g)$ for an asymmetric transition between $\gs$ and $\es$, which realizes an open quantum system without energy conservation.
The non-Hermiticity parameter $g$ promotes the transition to the excited state, which represents the situation in which the particle takes up energy from the environment.
We realized a system without momentum conservation by manipulating the parameter $\theta$ for the coin operator for a discrete-time quantum walk; we utilized the property that the continuum limit of a one-dimensional discrete-time quantum walk gives the Dirac equation with its mass proportional to the parameter $\theta$.
To realize dynamics under a harmonic potential in a two-dimensional system, we use the newly proposed unitary discrete-time quantum walk~\cite{Yamagishi23}, whose dynamics are similar to those of a Schr\"{o}dinger particle under a harmonic potential in two dimensions.

For our model, we have both found similarities to the previous research on a classical system~\cite{Schweitzer98} and observed unique quantum phenomena as follows:
\setlength{\leftmargin}{40pt}
	\setlength{\topsep}{0pt}
\begin{enumerate}
	\setlength{\itemsep}{0pt}      
	\setlength{\parskip}{0pt}      
	\setlength{\itemindent}{0pt}   
	\setlength{\labelsep}{5pt}     
  \renewcommand{\labelenumi}{(\Roman{enumi})}
  \item Similarities to classical active particles: The movement of the quantum walker becomes more active in a nontrivial manner as we increase the non-Hermiticity parameter $g$; examining the excited state and the ground state only, we can still clearly observe differences in the dynamics with different values of $g$.
  \item Unique quantum features: Oscillations emerge because of the resonant transition between the ground and excited states in one and two dimensions, the peak propagates ballistically in one dimension, and the quantum walker stays on a constant energy surface in two dimensions.
\end{enumerate}

Several open questions will be addressed in the future work.
While the reality of the quasienergy of the current model is protected by the pseudo-Hermiticity, constructing a $\mathcal{PT}$-symmetric model in a similar manner would provide an alternative approach for defining the quantum active particle.
It is interesting to study the symmetry classes of quantum active matter and compare them with those found in non-Hermitian physics~\cite{Kawabata19}.
Introducing decoherence will enable the direct comparison of the present quantum system to classical active Brownian systems.

Finally, we discuss the experimental realization of our model.
Quantum walks have already been realized in various systems, \textit{e.g.},\ cold atom systems~\cite{Mugel16}, laser systems~\cite{Schreiber10} and photon systems~\cite{Xiao17, Su19, Yan19}.
Focusing in particular on laser systems, we can realize the ground and excited states by utilizing optical devices with different transmittances.
By coupling the two devices with laser pumping, we may be able to realize our quantum active particle experimentally.
\textit{Note added: while we were finalizing our manuscript, we became aware of another paper on quantum active matter by Yuan et al.~\cite{Yuan24}}




%
%
%
%
%
%
%
%
%
%

\section*{Acknowledgments}
We are grateful to Ryo Hanai, Ken-Ichiro Imura, Kohei Kawabata, Kazue Kudo, Yuta Kuroda, Franco Nori, Kazuki Sone, Kazuaki Takasan and Kazumasa A.\ Takeuchi for fruitful discussions.
The computations in this work were performed partly using the facilities of the Supercomputer Center, the Institute for Solid State Physics, the University of Tokyo.
This work was supported by JSPS KAKENHI (Grant Numbers JP19H00658, JP21H01005, JP23K22411, JP24KJ0655 and JP24K00545).
M.Y.\ was supported by the RIKEN Junior Research Associate Program.

\section*{Author contributions statement}
M.Y.\ established the model based on the discussions with N.H.\ and H.O.\ and carried out all numerical calculations.
N.H.\ proposed to define a quantum version of active matter by using nonunitary quantum walks and call it a quantum active matter.
N.H.\ and H.O.\ subsequently joined the discussions with M.Y.
All authors reviewed the manuscript.

\section*{Competent interests}
The authors declare no competing interests.

\section*{Data availability}
The datasets used and/or analysed during the current study are available from the corresponding author on reasonable request.

\bibliography{yamagishi}

\end{document}


\title{Proposal of a quantum version of active particles via a nonunitary quantum walk:\\
Supplementary mateiral}
\author{Manami Yamagishi}
\email{manami@iis.u-tokyo.ac.jp}
\affiliation{Department of Physics, the University of Tokyo,
5-1-5 Kashiwanoha, Kashiwa, Chiba 277-8574, Japan}
\affiliation{Theoretical Quantum Physics Laboratory, RIKEN,
2-1 Hirosawa, Wako, Saitama 351-0198, Japan}
\author{Naomichi Hatano}
\email{hatano@iis.u-tokyo.ac.jp}
\affiliation{Institute of Industrial Science, the University of Tokyo,
5-1-5 Kashiwanoha, Kashiwa, Chiba 277-8574, Japan}
\author{Hideaki Obuse}
\email{hideaki.obuse@eng.hokudai.ac.jp}
\affiliation{Department of Applied Physics, Hokkaido University,
Kita13jo Nishi8cho, Kita, Sapporo, Hokkaido 060-8628, Japan}
\affiliation{Institute of Industrial Science, the University of Tokyo,
5-1-5 Kashiwanoha, Kashiwa, Chiba 277-8574, Japan}

\date{\today}

\begin{abstract}
\end{abstract}

\maketitle

\section{Review of quantum walks in one dimension and higher dimensions}
\label{appQW}

\subsection{One-dimensional quantum walks}

The quantum walk was first proposed in one dimension~\cite{Aharonov93}.
In addition to the spatial degree of freedom $\ket{x}$,
we let each lattice point have an internal degree of freedom of two states $\lk$ and $\rk$.
The entire state space is thereby spanned by $\ket{x}\otimes \lk$ and $\ket{x}\otimes\rk$.

The shift operator $\tilde{S}$ is defined to update the spatial part of the state according to the internal degree of freedom:
\begin{align}\label{appeqshift}
\tilde{S}&:=\sum_x\qty[\dyad{x-a}{x} \otimes \dyad{\mathrm{L}}
+\dyad{x+a}{x}\otimes \dyad{\mathrm{R}}],
\end{align}
where $a$ is the lattice constant.
In other words, the wave function of the component $\lk$ is shifted to the left by one lattice constant, while that of $\rk$ is shifted to the right by one lattice constant.
(Note that throughout the paper, we put the lattice constant $a$ to unity, for simplicity.)
If we kept operating the shift operator only, the left-going component would keep moving left, while the right-going component would keep moving right, without any interaction between them.
We hence additionally introduce the coin operator
\begin{align}\label{appeqcoin}
\tilde{C}&:=\sum_x\qty[\dyad{x}\otimes u],
\end{align}
where $u$ is a $2\times2$ unitary matrix that shuffles the left-going and right-going components at each lattice point.
Throughout the present paper we use the following specific form for the unitary $u$:
\begin{align}\label{appequ}
u=\mqty(\cos\theta & -\sin \theta \\
\sin\theta & \cos\theta).
\end{align}

\begin{figure}
  \centering
  \includegraphics[width=0.6\columnwidth]{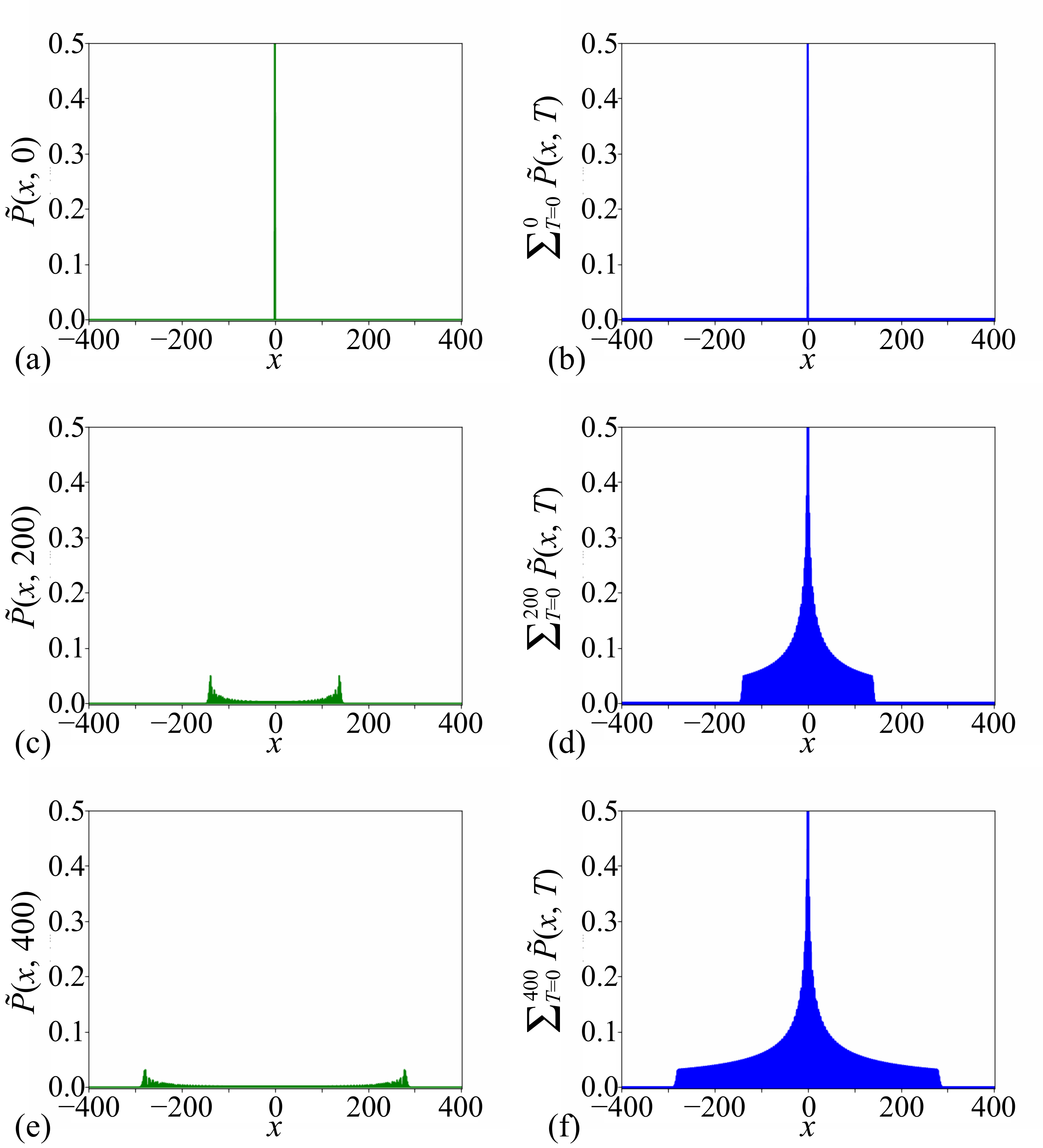}
  \caption{A numerical example of the probability distribution [(a), (c), (e)] and the temporal sum of the probability [(b), (d), (f)] of the quantum walker after $0$ [(a), (b)], $200$ [(c), (d)] and $400$ [(e), (f)] time steps of evolution in one dimension under the initial condition~\eqref{appeqinit}.}
  \label{appQWfig}
\end{figure}

Taking the initial state 
\begin{align}\label{appeqinit}
\ket{\psi(0)}=\frac{1}{\sqrt{2}}\ket{0}\otimes\qty(\lk+\ii\rk),
\end{align}
for example, and applying the shift and coin operators repeatedly as in
\begin{align}\label{appeqSC}
\ket{\psi(T)}=\qty(\tilde{S}\tilde{C})^T\ket{\psi(0)},
\end{align}
we observe the probability distribution $\braket{\psi(T)}$ in Fig.~\ref{appQWfig}(a).

We can understand this probability distribution as follows.
In classical random walks, the walker that starts from the origin takes various paths stochastically.
Each path has its own positive probability.
Since there are more paths around the starting point, the probability accumulates higher around there. 
In quantum walks, on the other hand, each path has its own amplitude, which is generally complex.
Although there are more paths around the starting point, they interfere with each other because of their random phases instead of the probability accumulation.
This interference decreases the probability around the origin.
Since the number of paths is small near the right and left wave fronts, the interference is weaker, and hence the probability has peaks there.

\subsection{Two-dimensional quantum walks}

A consistent definition of two-dimensional quantum walks was introduced in Ref.~\cite{Yamagishi23}. 
It starts with the identification of the one-dimensional Dirac Hamiltonian
\begin{align}\label{appeqHD1}
H_\mathrm{D}^\textrm{1d}=cp_x\sigma^z + m_x\sigma^y
\end{align}
as an approximate generator of the one-dimensional quantum walk~\cite{Strauch06},
where $c$ is the speed of light, $p_x$ is the momentum operator in the $x$ direction, $m_x$ is a mass term, and $\sigma^z$ and $\sigma^y$ are the Pauli matrices that represent the inner degree of freedom of the Dirac particle.
The Trotter decomposition of dynamics due to the Dirac Hamiltonian~\eqref{appeqHD1} yields the dynamics~\eqref{appeqSC},
where the light speed $c$ is given by the lattice constant divided by the unit time step and $m_x$ is proportional to $\theta$ in equation~\eqref{appequ}.

In squaring the Dirac Hamiltonian~\eqref{appeqHD1}, we note that the crossing term disappear because the Pauli matrices are elements of the two-dimensional Clifford algebra: $\sigma^z\sigma^y+\sigma^y\sigma^z=0$.
This motivates us to introduce~\cite{Yamagishi23} the Dirac Hamiltonian in two spatial dimensions in the form
\begin{align}\label{appeqHD2}
H_\mathrm{D}^\textrm{2d}=\qty(c p_x \sigma^z + m_x\sigma^y)\otimes\tau^0
+\sigma^z\otimes(c p_y \tau^z + m_y\tau^y),
\end{align}
where $p_y$ is the momentum operator in the $y$ direction, $m_y$ is an additional mass term, and $\tau^z$ and $\tau^y$ are the Pauli matrices in an additional inner degree of freedom with $\tau^0$ being the $2\times 2$ identity matrix in the additional space.
In squaring the newly introduced Dirac Hamiltonian~\eqref{appeqHD2} in two spatial dimensions, we note that all crossing terms disappear because the four operators $\sigma^z\otimes \tau^0$, $\sigma^y\otimes \tau^0$, $\sigma^x\otimes\tau^z$ and $\sigma^x\otimes\tau^y$ are elements of the four-dimensional Clifford algebra.
The Trotter decomposition of the dynamics due to the Dirac Hamiltonian~\eqref{appeqHD2} yields a consistent definition of quantum walk in two spatial dimensions~\cite{Yamagishi23}.
It is now straightforward to extend the quantum walk to three and higher-dimensional ones.

\section{Lack of orthonormality in non-Hermitian Hamiltonians}
\label{appendNHeigvec}
We here show that in spite of the fact that the pseudo-energy eigenvalues are all fixed to real values, because of the lack of orthonormality of the eigenvectors, the probability and the energy expectation value generally oscillate in time, as we stress at the end of the section defining the one-dimensional model in the main text. 

Let us consider a general non-Hermitian Hamiltonian $H$ with $H^{\dag}\neq H$.
Its eigenvectors are defined in the following manner:
\begin{align}\label{eqNH4}
\begin{cases}
H\ket{\psiR_n}=E_n\ket{\psiR_n} & \mbox{for a right-eigenvector $\ket{\psiR_n}$}, \\
\bra{\psiL_n}H=\bra{\psiL_n}E_n & \mbox{for a left-eigenvector $\bra{\psiL_n}$}.
\end{cases}
\end{align}
Note that the latter is often represented in the form
\begin{align}
H^\dag\ket{\psiL_n}={E_n}^\ast\ket{\psiL_n},
\end{align}
where $\ket{\psiL_n}:=\bra{\psiL_n}^\dag$.
In other words, the right- and left- eigenvectors of a non-Hermitian matrix $H$ are essentially the right-eigenvectors of $H$ and $H^\dag$, respectively.
With the definition~\eqref{eqNH4}, the eigenvectors satisfy bi-orthonormality
\begin{align}\label{eqNH9}
\ip{\psiL_n}{\psi_m^{\mathrm{R}}}=\delta_{nm}
\end{align}
under proper normalization, but the orthonormality is not in general satisfied:
\begin{align}\label{eqNH10}
\ip{\psiR_n}{\psi_m^{\mathrm{R}}}\neq\delta_{nm},
\end{align}
where $\bra{\psiR_n}:=\ket{\psiR_n}^\dag$.

Let us specifically look at our non-Hermitian Hamiltonian $H_{\NH}(g)$ in equation~(3) as an example. 
As shown in equation~(5), 
we can make the non-Hermitian matrix $H_{\NH}(g)$ Hermitian with the imaginary gauge transformation $A(g)$~\cite{Hatano96, Hatano97} 
with an imaginary vector potential $g$.
Let $\ket{\psi_n}$ denote the eigenvector for the Hermitian Hamiltonian $H_{\NH}(g=0)$:
\begin{align}
H_{\NH}(g=0)\ket{\psi_n}=E_n\ket{\psi_n}, \label{eqNH1} \\
\bra{\psi_n}H_{\NH}(g=0)=\bra{\psi_n}E_n. \label{eqNH2}
\end{align}
Note that in this case of the Hermitian Hamiltonian $H_\NH(g=0)$, the left-eigenvector $\bra{\psi_n}$ and the Hermitian conjugate of the right-eigenvector $\ket{\psi_n}$ are the same as in $\ket{\psi_n}=\bra{\psi_n}^\dag$ and the eigenvectors satisfy the standard orthonormality
\begin{align}\label{eqNH5}
\ip{\psi_n}{\psi_m}=\delta_{nm}.
\end{align}
From equations~(5) and~\eqref{eqNH1}, we obtain the following:
\begin{align}
A(g)^{-1}H_{\NH}(g)A(g)\ket{\psi_n}=E_n\ket{\psi_n}. \label{eqNH3}
\end{align}
Operating $A(g)$ from the left on each side of equation~\eqref{eqNH3} yields
\begin{align}\label{eqNH6}
H_{\NH}(g)[A(g)\ket{\psi_n}]=E_n[A(g)\ket{\psi_n}].
\end{align}
We thereby notice that if we take $\ket{\psiR_n}$ as in
\begin{align}\label{eqNH7}
\ket{\psiR_n}=A(g)\ket{\psi_n},
\end{align}
then equation~\eqref{eqNH6} is in the completely same form as the top line of equation~\eqref{eqNH4}.
Similarly, from equations~(5) and~\eqref{eqNH2}, we obtain
\begin{align}
[\bra{\psi_n}A(g)^{-1}]H_{\NH}(g)=[\bra{\psi_n}A(g)^{-1}]E_n.
\end{align}
Thus we take $\bra{\psiL_n}$ as in
\begin{align}\label{eqNH8}
\bra{\psiL_n}=\bra{\psi_n}A(g)^{-1}
\end{align}
to obtain the bottom line of equation~\eqref{eqNH4}.
From equations~\eqref{eqNH5}, \eqref{eqNH7} and \eqref{eqNH8}, we can easily check that the bi-orthonormality~\eqref{eqNH9} is satisfied while the orthonormality~\eqref{eqNH10} is not unless $A(g)^{\dag}A(g)=\mathbbm{I}$, which holds only for $g=0$ in our case of the non-Hermitian Hamiltonian $H_\NH(g)$.


Let us next consider the time evolution of eigenvectors of a general non-Hermitian Hamiltonian $H$.
We assume that the Schr\"{o}dinger equation for a non-Hermitian Hamiltonian $H$ has the forms
\begin{align}
\begin{cases}
\displaystyle
\ii \dv{t}\ket{\psiR(t)}=H\ket{\psiR(t)},\\
\displaystyle
-\ii \dv{t}\bra{\psiL(t)}=\bra{\psiL(t)}H,
\end{cases}
\end{align}
where we put $\hbar$ to unity.
The latter equation may look more plausible in the form
\begin{align}
\ii \dv{t}\ket{\psiL(t)}=H^\dag\ket{\psiL(t)}.
\end{align}
In other words, the right- and left-eigenvectors are driven by the non-Hermitian Hamiltonians $H$ and $H^\dag$, respectively.
This, combined with equation~\eqref{eqNH4}, results in the time-evolution of the forms 
\begin{align}
\begin{cases}
\ket{\psiR_n(T)}=\ee^{-\ii E_nT}\ket{\psiR_n(0)},\\
\bra{\psiL_n(T)}=\ee^{+\ii E_nT}\bra{\psiL_n(0)},
\end{cases}
\end{align}
which keeps the bi-orthonormality~\eqref{eqNH9} intact for any $T$.
Thus if we calculate the probability density with the left- and right-eigenvectors, it does not depend on time:
\begin{align}
\braket{\psiL(T)}{\psiR(T)}%
=\sum_{n,m}{c_m}^\ast c_n\ee^{\ii(E_m-E_n)T}\braket{\psiL_m}{\psiR_n}%
=\sum_n\abs{c_n}^2,
\end{align}
where we expanded the initial states as
\begin{align}
\begin{cases}
\ket{\psiR(0)}=\sum_n c_n\ket{\psiR_n},\\
\bra{\psiL(0)}=\sum_m {c_m}^\ast \bra{\psiL_m}.
\end{cases}
\end{align}
The energy expectation value defined in the form $\mel{\psiL(T)}{H}{\psiR(T)}$ is conserved because
\begin{align}\label{eq48}
\mel{\psiL(T)}{H}{\psiR(T)}%
=\sum_{n,m}{c_m}^\ast c_n\ee^{\ii(E_m-E_n)T}\mel{\psiL_m}{H}{\psiR_n}%
=\sum_n\abs{c_n}^2E_n.
\end{align}

On the other hand, if we calculate the probability density with the right-eigenvectors and their Hermitian conjugate, then it fluctuates depending on time as follows:
\begin{align}
\braket{\psiR(T)}{\psiR(T)}=\sum_{n,m}{c_m}^\ast c_n\ee^{\ii({E_m}^\ast-E_n)T}\braket{\psiR_m}{\psiR_n},
\end{align}
whose last factor is not reduced to $\delta_{mn}$ because of the inequality~\eqref{eqNH10}.
Note that since the imaginary gauge transformation $A(g)=\mathop{\mathrm{diag}}(\ee^{-g/2}, \ee^{-g/2}, \ee^{+g/2}, \ee^{+g/2})$ for our model is Hermitian, we have $A(g)^\dag A(g)=A(g)^2\neq \mathbb{I}$, and hence specifically 
\begin{align}
\braket{\psiR_m}{\psiR_n}=\mel{\psi_m}{A(g)^2}{\psi_n}\neq\delta_{mn}
\end{align}
unless $g=0$.
Similarly, we can check that the energy expectation value defined in the form $\mel{\psiR(T)}{H}{\psiR(T)}$ is not conserved because
\begin{align}\label{eq51}
\mel{\psiR(T)}{H}{\psiR(T)}%
=\sum_{n,m}{c_m}^\ast c_n\ee^{\ii({E_m}^\ast-E_n)T}\mel{\psiR_m}{H}{\psiR_n}
=\sum_{n,m}{c_m}^\ast c_n\ee^{\ii({E_m}^\ast-E_n)T}E_n\braket{\psiR_m}{\psiR_n}.
\end{align}

The reason why we use  in the present paper the right-eigenvectors and their Hermitian conjugate is because our model is an open quantum system. There are two ways of understanding non-Hermitian systems; one is the closed non-Hermitian system and the other is the open quantum system~\cite{Kawabata19,Hatano21}. The original concept of the $PT$-symmetric non-Hermitian system~\cite{Bender98,Bender02} was to replace the Hermiticity with more physical symmetries in making certain that the expectation values of the energy and other physical observables are real. Since the system is considered to be closed in this framework, we should employ the right- and left-eigenvectors to keep the bi-orthogonality~\eqref{eqNH9} satisfied and to make the energy expectation value constant in time as in equation~\eqref{eq48}. 

The open quantum system, on the other hand, is considered to be embedded in an environment and the entire system of the central system combined with the environment is supposedly Hermitian. Therefore, we should employ the standard way of calculating the expectation value of an observable, if it is localized in the central system or not, by employing the right-eigenvectors and their Hermitian conjugate as in equation~\eqref{eq51}. Since the open quantum system can exchange the probability and the energy with the environment, it is natural that the probability density and the energy expectation value are not constant in time (cf.~\cite{Makris08,Ruter10,Hatano24}).

\section{Peak velocity of the quantum walkers}\label{appvg}

We here present the derivation of the peak velocity of quantum walkers, which we used in the analyses of Fig.~3.
Consider first the case without the operator $N(g)$, for which the walkers in the ground and excited states are independent of each other.
We can then derive the peak velocity as follows.
For brevity, let us drop the suffices G and E for the moment and put $\hbar=a=1$, so that $x$ is an integer.

The walker in each state evolves in time according to the following shift and coin operators:
\begin{align}
\tilde{S}:=\sum_x\qty[\dyad{x-1}{x} \otimes 
\mqty( 1 & 0 \\ 0 & 0)
+\dyad{x+1}{x} \otimes
\mqty(0 & 0 \\ 0 & 1)],
\quad
\tilde{C}:=\sum_x\qty[\dyad{x}\otimes 
\mqty( \cos\theta & -\sin\theta \\
\sin\theta & \cos\theta)],
\end{align}
where we represent the inner degree of freedom of two states $\lk$ and $\rk$ in equations~\eqref{appeqshift}--\eqref{appequ}  in terms of $2\times2$ matrices and we put the lattice constant $a$ to unity.
In order to analyze the dynamics efficiently,
let us introduce the Fourier transformation in the following forms:
\begin{align}
\ket{x}=\frac{1}{\sqrt{2\pi}}\int_{-\pi}^{\pi}e^{-\ii k x}\ket{k}\dd k,
\quad
\ket{k}=\frac{1}{\sqrt{2\pi}}\sum_{x=-\infty}^\infty e^{\ii k x}\ket{x}.
\end{align}
It casts the shift and coin operators into the forms
\begin{align}
\tilde{S}=\int_{-\pi}^\pi \dd k\dyad{k}\otimes 
\mqty( \ee^{\ii k} & 0 \\ 0 & \ee^{-\ii k}),
\quad
\tilde{C}=\int_{-\pi}^\pi \dd k\dyad{k}\otimes 
\mqty( \cos\theta & -\sin\theta \\
\sin\theta & \cos\theta).
\end{align}
We are now in a position to analyze each subspace of $k$ separately.
The two eigenvalues of 
\begin{align}
\mqty( \ee^{\ii k} & 0 \\ 0 & \ee^{-\ii k})
\mqty( \cos\theta & -\sin\theta \\
\sin\theta & \cos\theta)
\end{align}
are given in the form of
\begin{align}
\lambda_\pm(k):=\cos k \cos\theta\pm\ii\sqrt{1-\cos^2 k \cos^2\theta}
\end{align}
or
\begin{align}
\lambda_\pm(k)=\ee^{\pm\ii E(k)}
\end{align}
with the relation~\cite{Kempf09}
\begin{align}\label{eqdisp}
\cos E(k)=\cos k \cos\theta.
\end{align}

The last relation gives the dispersion relation of the quantum walk.
The group velocity is then given by
\begin{align}\label{eqvgk}
v_\textrm{g}(k):=\dv{E}{k}=\frac{\sin k}{\sin E(k)}\cos\theta.
\end{align}
Its maximum with respect to $k$ should give the peak velocity of the quantum walk because the peaks are the front runners.
By further differentiating equation~\eqref{eqvgk}, we have
\begin{align}
\dv{v_\textrm{g}}{k}=\qty(\frac{\cos k}{\sin E}-\frac{\sin k \cos E\times v_\textrm{g}}{\sin^2 E})\cos\theta%
=\frac{\cos k}{\sin^3 E}\sin^2\theta\cos\theta.
\end{align}
This thereby shows that the component of $k=\pi/2$ runs fastest and its group velocity gives the peak velocity of the free random walker.
Since equation~\eqref{eqdisp} produces $\cos E(k)=0$ and $\sin E(k)=\pm1$ at $k=\pi/2$, the peak velocity of the free random walker is $v_\textrm{g}(\pi/2)=\cos\theta$.
We used this fact when we fixed as
$\cos\theta_\mathrm{g}\simeq 0.270$ and
$\cos\theta_\mathrm{e}\simeq 0.698$ from Fig.~3 in the main text.

\begin{figure}[h!]
  \centering
  \includegraphics[width=0.75\columnwidth]{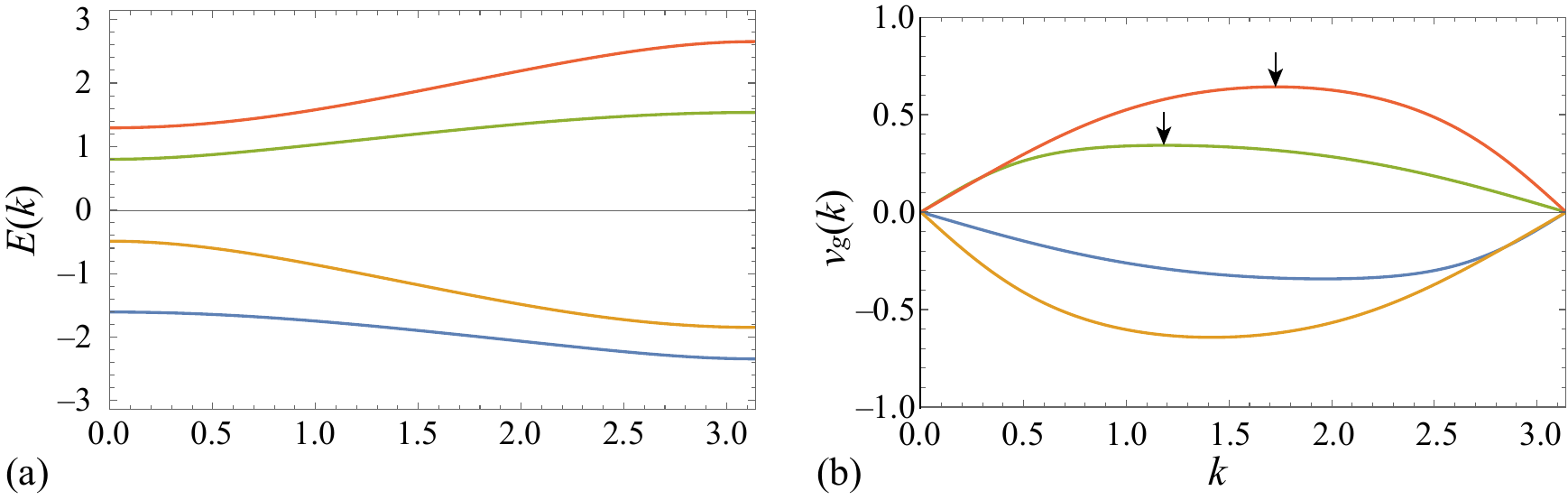}
  \caption{(a) The $k$-dependence of the four energy eigenvalues; in other words, the dispersion relation.
  (b) The $k$-dependence of the group velocity. The arrows indicate the maxima of the group velocities for the first and second largest eigenvalues.
  The parameter values are the ones specified in  equation~(15).
  We do not show the values for $-\pi<k<0$ because the dispersion relation is symmetric with respect to $k=0$ and the group velocity is $\pi$-periodic.}
  \label{figB}
\end{figure}

When we have a finite operator $N(g)$, which connects the ground and excited states, we should move over to the $4\times 4$ matrices, again in the Fourier space:
\begin{align}
C_k&:=\mqty(
\cos\theta_\textrm{g} & -\sin\theta_\textrm{g} & & \\
\sin\theta_\textrm{g} & \cos\theta_\textrm{g} & & \\
& & \cos\theta_\textrm{e} & -\sin\theta_\textrm{e}  \\
& & \sin\theta_\textrm{e} & \cos\theta_\textrm{e} \\
),\label{eqCk}
\\
S_k&:=\mqty(
\ee^{\ii k} &&&\\
&\ee^{-\ii k} &&\\
&&\ee^{\ii k}&\\
&&&\ee^{-\ii k}
),\label{eqSk}
\\
N_k(g)&:=\exp\qty[-\ii\mqty(
-\varepsilon &  & -w\ee^{-g} & \\
& -\varepsilon & & -w\ee^{-g} \\
-w\ee^{+g} & & \varepsilon & \\
& -w\ee^{+g} & & \varepsilon)]%
=\mqty(
\xi-\ii\varepsilon \eta & & \ii w \ee^{-g} \eta & \\
& \xi-\ii\varepsilon\eta & & \ii w \ee^{-g}\eta \\
\ii w\ee^{+g}\eta & & \xi+\ii \varepsilon \eta &\\
& \ii w\ee^{+g}\eta & & \xi+\ii \varepsilon \eta 
),\label{eqNk}
\end{align}
where 
\begin{align}
\xi:=\cos\sqrt{\varepsilon^2+w^2},
\quad
\eta:=\frac{\sin\sqrt{\varepsilon^2+w^2}}{\sqrt{\varepsilon^2+w^2}}.
\end{align}
We thereby numerically find the four eigenvalues of $\ii \ln S_kC_kN_k(g)$ in the case of equation~(15) as in Fig.~\ref{figB}(a) and compute the group velocity according to each eigenvalue as in Fig.~\ref{figB}(b). 
For $\varepsilon=w=0$, the red and green curves as well as the orange and blue curves in Fig.~\ref{figB} would be degenerate.
Introduction of finite values of $w$ and $\varepsilon$ lift the degeneracy;
now the green and blue curves are for the ground state, while the red and orange curves are for the excited states. 
We can indeed observe that the corresponding group velocity for the excited state is greater than the one for the ground state for all $k$ except for $k=0$ and $k=\pi$.
This numerically confirms the expectation that the quantum walker runs faster in the excited state  than in the ground state.

For the specific parameter values of equation~(15), the maximum of each group velocity  gives about $\pm 0.343$ at $k\simeq 0.624\pi$ and $\pm0.642$ at $k\simeq 0.450\pi$ as indicated by arrows in the figure, which we quoted in the analyses of Fig.~3.

\section{Probability oscillations between the ground and excited states}
\label{apposc}



Here we explain that our active quantum walk  is approximately described by an effective two-level Hamiltonian.
The description is  consistent with the  oscillations found in Fig.~4. 

\begin{figure}
\centering
\includegraphics[width=0.38\columnwidth]{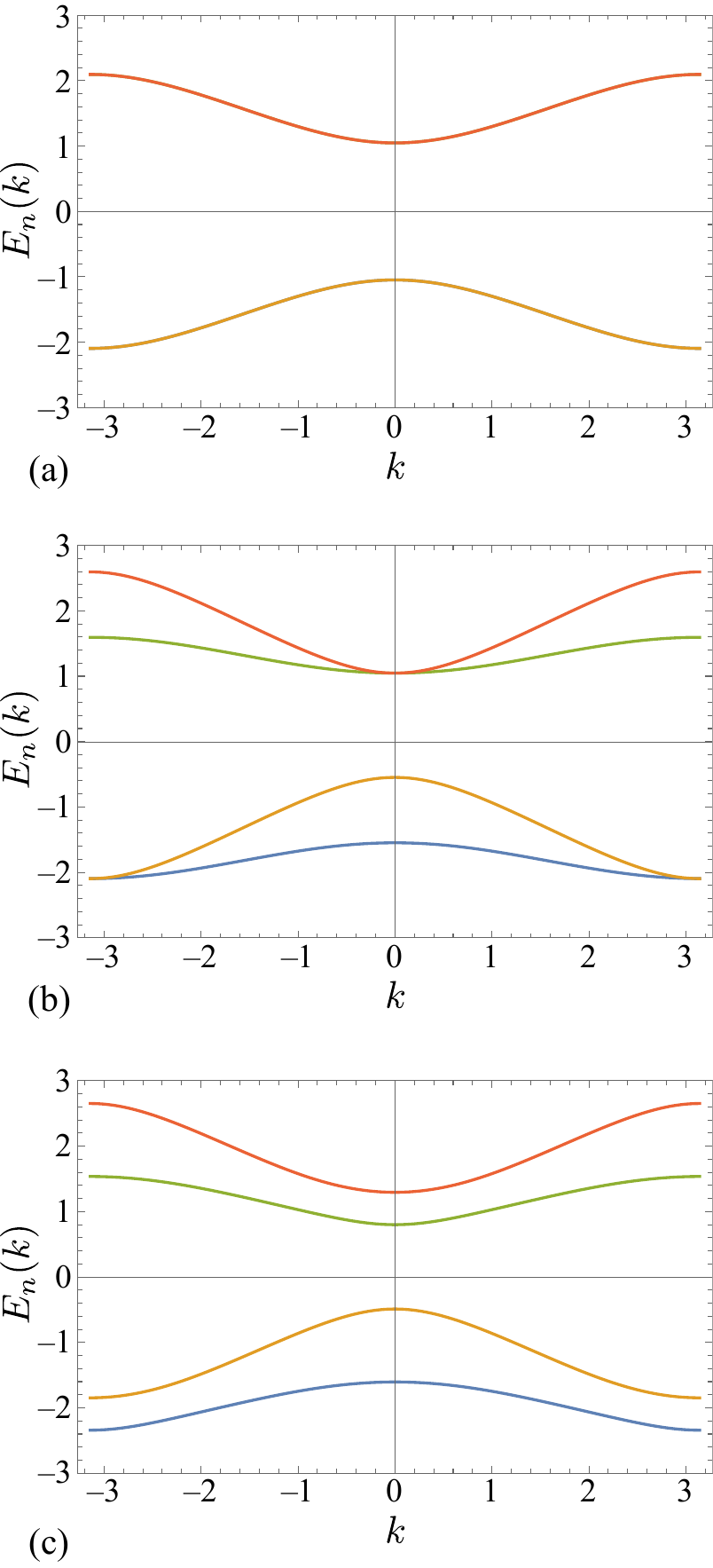}
\caption{The dispersion relation of the four pseudo-energy eigenvalues for (a) $\varepsilon=w=0$, (b) $\varepsilon=0.25$ with $w=0$, and (c) $\varepsilon=w=0.25$. In Panel (a), two curves are degenerate in each of the two curves. Panel (c) is equivalent to Fig.~\ref{figB}(a) but in a doubled plotting region. In all cases, we set $\theta_0=\pi/3$.}
\label{figD}
\end{figure}

%
%
We notice in Fig.~\ref{figB}(a), or equivalently in Fig.~\ref{figD}(c) that the upper two levels are the closest to each other at $k=0$, and symmetrically, the lower two levels at $k=\pi$.
We thereby approximately derive an effective Hamiltonian for each set of two levels and show that resonant oscillations between the two levels at the minimum energy gap is quantitively consistent with the oscillation that we find in Fig.~4. 
We also show an amplification of the probability in the excited level by the factor of $\ee^{2g}$, which is also consistent with equation~(10).

As shown in Fig.~\ref{figD}(a), the ground and excited states are degenerate for $\varepsilon=w=0$, for which the upper and lower blocks of $C_k$ in equation~\eqref{eqCk} are equal to each other and $N_k(g)$ in equation~\eqref{eqNk} is reduced to the identity operator.
When we turn on $\varepsilon$ to $0.25$ but keep $w=0$ as in Fig.~\ref{figD}(b), the degeneracy between the ground and excited states are lifted except for $k=0$ for the upper levels and $k=\pi$ for the lower levels.
When we further turn on $w$ to $0.25$ as in Fig.~\ref{figD}(c), the degeneracies at $k=0$ and $k=\pi$ are both lifted.
We claim that this smallest energy gaps cause the oscillations found in Fig.~4. 

Let us here focus on the energy gap at $k=0$ between the upper two levels, and estimate it by perturbation theory up to the first order of $w$, keeping $\varepsilon$ finite.
For $k=0$, the shift operator~\eqref{eqSk} is reduced to the identity operator, and hence irrelevant. The coin operator~\eqref{eqSk} keeps the same form.
For the operator~\eqref{eqNk}, the unperturbed limit yields
\begin{align}
N_g^{(0)}=\begin{pmatrix}
\ee^{\ii\varepsilon} & & & \\
&\ee^{\ii\varepsilon} & & \\
& & \ee^{-\ii\varepsilon} & \\
& & & \ee^{-\ii\varepsilon} 
\end{pmatrix},
\end{align}
whereas the first-order perturbation is given by
\begin{align}
N_g^{(1)}=\ii w \eta_0 
\begin{pmatrix}
& & \ee^{-g} & \\
& & & \ee^{-g}\\
\ee^{+g} & & & \\
& \ee^{+g} & & 
\end{pmatrix}
\end{align}
with $\eta_0=\sin (\varepsilon)/\varepsilon$.
The perturbation parameter $w$ in $\xi$ and $\eta$ of equation~(C16) only contributes to the second-order perturbation, which we neglect hereafter.

To summarize, the unperturbed time-evolution operator for $k=0$ is
\begin{align}
U_0^{(0)}&:=C_0N_g^{(0)}
\notag\\
&=
\begin{pmatrix}
\ee^{\ii\varepsilon}
\begin{pmatrix}
\cos\theta_\mathrm{g} & -\sin\theta_\mathrm{g} \\
\sin\theta_\mathrm{g} & \cos\theta_\mathrm{g} 
\end{pmatrix}
& \\
& \ee^{-\ii\varepsilon}
\begin{pmatrix}
\cos\theta_\mathrm{e} & -\sin\theta_\mathrm{e}  \\
\sin\theta_\mathrm{e} & \cos\theta_\mathrm{e} 
\end{pmatrix} 
\\
\end{pmatrix}
=\begin{pmatrix}
\ee^{-\ii(\theta_g\sigma_y-\varepsilon)} & \\
& \ee^{-\ii(\theta_e\sigma_y+\varepsilon)}
\end{pmatrix},
\end{align}
while the first-order perturbation is given by
\begin{align}
\label{eqU0}
&U_0^{(1)}:=C_0N_g^{(1)}%
=\ii w \eta_0 
\begin{pmatrix}
& 
\ee^{-g}
\begin{pmatrix}
\cos\theta_\mathrm{g} & -\sin\theta_\mathrm{g} \\
\sin\theta_\mathrm{g} & \cos\theta_\mathrm{g} 
\end{pmatrix} \\
\ee^{+g}
\begin{pmatrix}
\cos\theta_\mathrm{e} & -\sin\theta_\mathrm{e}  \\
\sin\theta_\mathrm{e} & \cos\theta_\mathrm{e} 
\end{pmatrix}
& 
\end{pmatrix}.
\end{align}
The unperturbed eigenvectors that diagonalize the unperturbed matrix~\eqref{eqU0} are 
\begin{align}
\label{eqeigvec0}
&\ket{\psi^{(0)}_{\mathrm{g}+}}=\frac{1}{\sqrt{2}}
\begin{pmatrix}
1 \\ \ii \\ 0 \\ 0
\end{pmatrix},
\quad
\ket{\psi^{(0)}_{\mathrm{e}+}}=\frac{1}{\sqrt{2}}
\begin{pmatrix}
0 \\ 0 \\ 1 \\ \ii 
\end{pmatrix},
\quad
\ket{\psi^{(0)}_{\mathrm{g}-}}=\frac{1}{\sqrt{2}}
\begin{pmatrix}
1 \\ -\ii \\ 0 \\ 0
\end{pmatrix},
\quad
\ket{\psi^{(0)}_{\mathrm{e}-}}=\frac{1}{\sqrt{2}}
\begin{pmatrix}
0 \\ 0 \\1 \\ -\ii
\end{pmatrix},
\end{align}
which give the unperturbed eigenvalues as
\begin{align}
\label{eqeigval0}
\lambda^{(0)}_{\mathrm{g}+}=\ee^{-\ii(\theta_\mathrm{g}-\varepsilon)}=\ee^{-\ii\theta_0},
\quad
\lambda^{(0)}_{\mathrm{e}+}=\ee^{-\ii(\theta_\mathrm{e}+\varepsilon)}=\ee^{-\ii\theta_0},
\quad
\lambda^{(0)}_{\mathrm{g}-}=\ee^{-\ii(-\theta_\mathrm{g}-\varepsilon)},
\quad
\lambda^{(0)}_{\mathrm{e}-}=\ee^{-\ii(-\theta_\mathrm{e}+\varepsilon)}.
\end{align}
The eigenvalues $\lambda^{(0)}_{\mathrm{g}+}$ and $\lambda^{(0)}_{\mathrm{e}+}$ in equation~\eqref{eqeigval0} correspond to the positive pseudo-energies, which are degenerate to $\theta_0$ as in the upper two levels of Fig.~\ref{figD}(b).
(The eigenvalues $\lambda^{(0)}_{\mathrm{g}-}$ and $\lambda^{(0)}_{\mathrm{e}-}$ are degenerate to the negative pseudo-energy $-\theta_0$ at $k=\pi$ rather than $k=0$.)

In other words,  the unperturbed eigenvectors 
in equation~\eqref{eqeigvec0} both give the same eigenvalue $\ee^{-\ii\theta_0}$.
We therefore diagonalized the following matrix to find the first-order perturbed eigenvalues:
\begin{align}
\Delta U&=\begin{pmatrix}
\mel{\psi^{(0)}_{\mathrm{g}+}}{U_0^{(1)}}{\psi^{(0)}_{\mathrm{g}+}} &
\mel{\psi^{(0)}_{\mathrm{g}+}}{U_0^{(1)}}{\psi^{(0)}_{\mathrm{e}+}} \\
&\\
\mel{\psi^{(0)}_{\mathrm{e}+}}{U_0^{(1)}}{\psi^{(0)}_{\mathrm{g}+}} &
\mel{\psi^{(0)}_{\mathrm{e}+}}{U_0^{(1)}}{\psi^{(0)}_{\mathrm{e}+}} 
\end{pmatrix}.
\end{align}
Straightforward algebra produces
\begin{align}
\Delta U&=\ii w\eta_0\ee^{-\ii\theta_0} 
\begin{pmatrix}
0 & \ee^{-g-\ii\varepsilon} \\ 
\ee^{+g+\ii\varepsilon} & 0
\end{pmatrix}.
\end{align}
Since the eigenvalues of the matrix above are $\pm 1$, we have the eigenvalues  up to the first order of $w$ in the form
\begin{align}
\lambda_\pm \simeq \ee^{-\ii\theta_0}
\left(1\pm \ii w\eta_0\right).
\end{align}
The pseudo-energy eigenvalues are thereby given by
\begin{align}
E_\pm:=\ii\log\lambda_\pm\simeq\theta_0\mp w\eta_0.
\end{align}
Therefore, the energy gap at $k=0$ is found to be $\Delta E=2w\eta_0$.
For $\varepsilon=0.25$ and $w=0.25$, we have $\Delta E\simeq0.494808$.
The period of the oscillation is $2\pi/\Delta E\simeq12.6982$, which is consistent with the oscillation found in Fig.~4. 

Let us finally find the effective Hamiltonian that governs the dynamics in the Hilbert subspace of 
the two vectors in equation~\eqref{eqeigvec0}.
The matrix 
\begin{align}
V:=
\begin{pmatrix}
1 & \ee^{-g-\ii\varepsilon} \\
\ee^{+g+\ii\varepsilon} & -1
\end{pmatrix}
\end{align}
 diagonalizes $\Delta U$ with
\begin{align}
V^{-1}=\frac{1}{-2}\begin{pmatrix}
-1 & -\ee^{-g-\ii\varepsilon} \\
-\ee^{+g+\ii\varepsilon} & 1
\end{pmatrix}=\frac{V}{2}
\end{align}
as in $V^{-1}\Delta UV=\mathop{\textrm{diag}}\left(\lambda_+,\lambda_-\right)$.
This implies that the effective Hamiltonian is given by
\begin{align}
H_\textrm{eff}:=V\begin{pmatrix}
E_+ & \\
& E_- 
\end{pmatrix}
V^{-1}%
=\begin{pmatrix}
\theta_0 & -w\eta_0\ee^{-g-\ii\varepsilon} \\
-w\eta_0\ee^{+g+\ii\varepsilon} & \theta_0
\end{pmatrix}.
\end{align}
The effective time-evolution operator within this Hilbert subspace is therefore found to be
\begin{align}
{U_\textrm{eff}}^T:=\ee^{-\ii H_\textrm{eff} T}
=\ee^{-\ii\theta_0T}\exp\left[\ii w\eta_0T
\begin{pmatrix}
0 & \ee^{-g-\ii\varepsilon} \\
\ee^{+g+\ii\varepsilon} &0
\end{pmatrix}
\right]
\end{align}

In calculating the matrix exponential above, we take full advantage of a complex version of the gauge transformation  in equation~(5). Using the  gauge transformation
\begin{align}
A(g+\ii\varepsilon)=\begin{pmatrix}
\ee^{-(g+\ii\varepsilon)/2} & 0 \\
0 & \ee^{+(g+\ii\varepsilon)/2}
\end{pmatrix},
\end{align}
we have
\begin{align}
A^{-1}
\begin{pmatrix}
0 & \ee^{-g-\ii\varepsilon} \\
\ee^{+g+\ii\varepsilon}  &0
\end{pmatrix}
A=\begin{pmatrix}
0 & 1 \\
1 & 0
\end{pmatrix}.
\end{align}
We therefore find
\begin{align}
\exp\left[\ii w\eta_0T
\begin{pmatrix}
0 & \ee^{-g-\ii\varepsilon} \\
\ee^{+g+\ii\varepsilon} &0
\end{pmatrix}
\right]%
=A
\begin{pmatrix}
\cos(w\eta_0T) & \ii\sin(w\eta_0 T) \\
\ii\sin(w\eta_0 T) &\cos(w\eta_0T) 
\end{pmatrix}
A^{-1}%
=\begin{pmatrix}
\cos(w\eta_0T) & \ii\ee^{-g-\ii\varepsilon}\sin(w\eta_0 T) \\
\ii\ee^{+g+\ii\varepsilon}\sin(w\eta_0T) &\cos(w\eta_0T) 
\end{pmatrix}
\end{align}

\begin{figure}
\centering
\includegraphics[width=0.48\columnwidth]{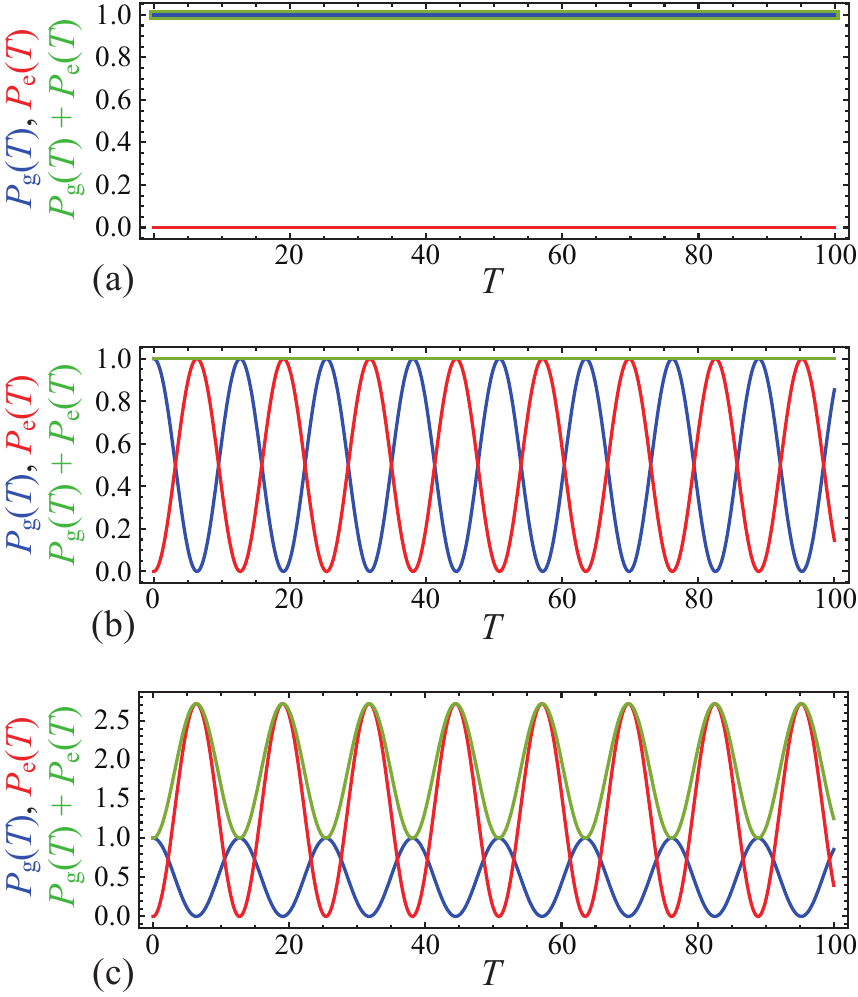}
\caption{The oscillation of the probabilities in equations~\eqref{proboscillation1} and \eqref{proboscillation2}. 
In each panel, the horizontal axis indicates $t$,  the blue curve indicates $ P_\mathrm{g}(t)$, the red curve indicates  $P_\mathrm{e}(t) $, and the green curve indicates the summation.
(a) $w=0$ and $g=0$. (b) $w=0.25$ and $g=0$. (c) $w=0.25$ and $g=0.5$.}
\label{fig20}
\end{figure}

Suppose, for example, that we start the time evolution from $\ket{\psi^{(0)}_{\mathrm{g}+}}$. 
We thereby find
\begin{align}
\ket{\psi_+(T)}
:={U_\textrm{eff}}^T\ket{\psi^{(0)}_{\mathrm{g}+}}
=e^{-\ii\theta_0 T}
\begin{pmatrix}
\cos(w\eta_0T)  \\
\ii\ee^{+g+\ii\varepsilon}\sin(w\eta_0 T) 
\end{pmatrix}
\end{align}
Therefore the following probabilities for the ground and excited states oscillate as in
\begin{align}\label{proboscillation1}
P_\textrm{g}(T)&:=\abs{\mel{\psi^{(0)}_{\mathrm{g}+}}{{U_\textrm{eff}}^T}{\psi^{(0)}_{\mathrm{g}+}}}^2=\cos^2(w\eta_0T) ,\\
\label{proboscillation2}
P_\textrm{e}(T)&:=\abs{\mel{\psi^{(0)}_{\mathrm{e}+}}{{U_\textrm{eff}}^T}{\psi^{(0)}_{\mathrm{g}+}}}^2=\ee^{2g}\sin^2(w\eta_0 T);
\end{align}
%
see Fig.~\ref{fig20}.
For $g=0$, this simply describes the resonant oscillation between the ground and excited states with the probability conservation $P_\mathrm{g}(T) + P_\mathrm{e}(T)=1$.
For $g>0$, we find the enhancement of the oscillation amplitude of the excited state by the factor of $\ee^{2g}$, which is consistent with equation~(10).
The oscillation should describe the behavior that we observe in Fig.~4.

For $w=0$ as in Fig.~\ref{fig20}(a), the ground and excited states of the effective Hamiltonian are degenerate and we do not observe any difference or oscillation between the two states.
This corresponds to the case of Fig.~4(a).
As we introduce the off-diagonal elements of $N_g$ to be $w=0.25$, the oscillation of the phase difference of $\pi$ for the ground and excited states appears with the period $\pi/(w\eta_0)\simeq 12.6982$ for $\varepsilon=0.25$, or about 8 oscillations in the range $0\leq T\leq 100$, which is consistent with what we observe in Fig.~4(b).
We notice nonetheless that the value for the sum of the two states (green line) constantly increases without any oscillations in Fig.~4(b). This corresponds to the probability conservation $P_\mathrm{g}(T) + P_\mathrm{e}(T)=1$.
Finally, when we introduce $g$ as in Fig.~4(c) and (d), we see that the green curve now oscillates with the same phase as the value for the excited state.
This implies that the oscillation amplitude of the excited state is enhanced by the factor $\ee^{2g}$, which is consistent with the observation in equation~\eqref{proboscillation2}.

\section{Quantum active particle in two dimensions}\label{2Dmodel}

One of our motivations of defining a quantum active particle is to find similarities and differences between our quantum mode with classical active particle models.
A good reference point of the classical active particle is provided by Ref.~\cite{Schweitzer98}, which observed active dynamics in two dimensions.
We here define a two-dimensional version of our quantum active particle for the purpose of comparing it with the model in Ref.~\cite{Schweitzer98},

\subsection{Definition of the two-dimensional quantum active particle}

We define a quantum active particle in two dimensions, making the discrete-time quantum walk in two dimensions proposed in Ref.~\cite{Yamagishi23} non-unitary.
Our quantum active walker now has eight internal states in total, namely, $(\lk\oplus\rk)\otimes(\dk\oplus\uk)\otimes(\gs\oplus\es)=\ket{\mathrm{LDG}}\oplus\ket{\mathrm{RDG}}\oplus\ket{\mathrm{LUG}}\oplus\ket{\mathrm{RUG}}\oplus\ket{\mathrm{LDE}}\oplus\ket{\mathrm{RDE}}\oplus\ket{\mathrm{LUE}}\oplus\ket{\mathrm{RUE}}$.
Here, $\uk$, $\dk$, $\rk$ and $\lk$ denote upward, downward, rightward and leftward states, respectively, while $\ket{\mathrm{G}}$ and $\ket{\mathrm{E}}$ denote the ground and excited states, respectively.
We  fix the ordering of the basis vectors in this way in the present section. 
We define the time evolution of our two-dimensional quantum active particle $\ket{\psi(T)}=[{U^\textrm{2d}(g)}]^T\ket{\psi(0)}$ for $T\in\mathbb{Z}$ in terms of the following operators:
\begin{align}
N^\textrm{2d}(g)&:=\sum_{x, y}\qty[
\dyad{x, y}{x, y}\otimes \ee^{-\ii H_{\NH}^\textrm{2d}(g)}
], \label{eqQABP20} \\
C_x^\textrm{2d}&:=\displaystyle\bigotimes_{x, y}\ee^{-\ii\theta_{x, \mathrm{G}}(x)(\sigma^y\otimes\tau^0\otimes\upsilon^{\mathrm{G}})-\ii\theta_{x, \mathrm{E}}(x)(\sigma^y\otimes\tau^0\otimes\upsilon^{\mathrm{E}})}
=\sum_{x, y}\qty[
\dyad{x, y}{x, y}\otimes C_x^\textrm{2d}(x)
], \label{eqQABP21} \\
S_x^\textrm{2d}&:=\ee^{-a(\sigma^z\otimes\tau^0\otimes\upsilon^0)\partial_x}
=\sum_{x, y}\qty[
\dyad{x, y}{x, y}\otimes S_x^\textrm{2d}(x)
], \label{eqQABP22} \\
C_y^\textrm{2d}&:=\displaystyle\bigotimes_{x, y}\ee^{-\ii\theta_{y, \mathrm{G}}(y)(\sigma^x\otimes\tau^y\otimes\upsilon^{\mathrm{G}})-\ii\theta_{y, \mathrm{E}}(y)(\sigma^x\otimes\tau^y\otimes\upsilon^{\mathrm{E}})}
=\sum_{x, y}\qty[
\dyad{x, y}{x, y}\otimes C_y^\textrm{2d}(y)
], \label{eqQABP23} \\
S_y^\textrm{2d}&:=\ee^{-a(\sigma^x\otimes\tau^z\otimes\upsilon^0)\partial_y}
=\sum_{x, y}\qty[
\dyad{x, y}{x, y}\otimes S_y^\textrm{2d}(y)
], \label{eqQABP24}
\end{align}
with $U^\textrm{2d}(g):=S_y^\textrm{2d}C_y^\textrm{2d}S_x^\textrm{2d}C_x^\textrm{2d}N^\textrm{2d}(g)$. Here,
\begin{align}
H_{\NH}^\textrm{2d}(g):=\sigma^0\otimes\tau^0\otimes\mqty(
-\varepsilon & -w \ee^{-g}  \\
-w \ee^{+g} & +\varepsilon
).\label{eqQABP20-1}
\end{align}
The operators $C_x^\textrm{2d}(x)$, $S_x^\textrm{2d}(x)$, $C_y^\textrm{2d}(y)$ and $S_y^\textrm{2d}(y)$ read\newpage
\begin{align}
&C_x^\textrm{2d}(x)\notag\\%
&=\mqty(
+c_{x, \mathrm{G}}(x) & -s_{x, \mathrm{G}}(x) & & \\
+s_{x, \mathrm{G}}(x) & +c_{x, \mathrm{G}}(x) & & \\
 & & +c_{x, \mathrm{G}}(x) & -s_{x, \mathrm{G}}(x) \\
 & & +s_{x, \mathrm{G}}(x) & +c_{x, \mathrm{G}}(x)
)\otimes\upsilon^{\mathrm{G}}
+\mqty(
+c_{x, \mathrm{E}}(x) & -s_{x, \mathrm{E}}(x) & & \\
+s_{x, \mathrm{E}}(x) & +c_{x, \mathrm{E}}(x) & & \\
 & & +c_{x, \mathrm{E}}(x) & -s_{x, \mathrm{E}}(x) \\
 & & +s_{x, \mathrm{E}}(x) & +c_{x, \mathrm{E}}(x)
)\otimes\upsilon^{\mathrm{E}}, \\
&S_x^\textrm{2d}(x)%
=\mqty(
P_x & & & \\
 & Q_x & & \\
 & & P_x & \\
 & & & Q_x
)\otimes\upsilon^0, \\
&C_y^\textrm{2d}(y)\notag\\%
&=\mqty(
+c_{y, \mathrm{G}}(y) & & & -s_{y, \mathrm{G}}(y) \\
 & +c_{y, \mathrm{G}}(y) & -s_{y, \mathrm{G}}(y) & \\
 & +s_{y, \mathrm{G}}(y) & +c_{y, \mathrm{G}}(y) & \\
+s_{y, \mathrm{G}}(y) & & & +c_{y, \mathrm{G}}(y)
)\otimes\upsilon^{\mathrm{G}}
+\mqty(
+c_{y, \mathrm{E}}(y) & & & -s_{y, \mathrm{E}}(y) \\
 & +c_{y, \mathrm{E}}(y) & -s_{y, \mathrm{E}}(y) & \\
 & +s_{y, \mathrm{E}}(y) & +c_{y, \mathrm{E}}(y) & \\
+s_{y, \mathrm{E}}(y) & & & +c_{y, \mathrm{E}}(y)
)\otimes\upsilon^{\mathrm{E}}, \\
&S_y^\textrm{2d}(y)%
=\mqty(
P_y & Q_y & & \\
Q_y & P_y & & \\
 & & P_y & -Q_y \\
 & & -Q_y & P_y
)\otimes\upsilon^0,
\end{align}
where
\begin{align} 
c_{x, \mathrm{G}}(x):=\cos{(\theta_{x, \mathrm{G}}(x))},\quad%
s_{x, \mathrm{G}}(x):=\sin{(\theta_{x, \mathrm{G}}(x))},\quad%
&c_{x, \mathrm{E}}(x):=\cos{(\theta_{x, \mathrm{E}}(x))},\quad%
s_{x, \mathrm{E}}(x):=\sin{(\theta_{x, \mathrm{E}}(x))}, \\[2pt]
c_{y, \mathrm{G}}(y):=\cos{(\theta_{y, \mathrm{G}}(y))},\quad%
s_{y, \mathrm{G}}(y):=\sin{(\theta_{y, \mathrm{G}}(y))}, \quad%
&c_{y, \mathrm{E}}(y):=\cos{(\theta_{y, \mathrm{E}}(y))},\quad%
s_{y, \mathrm{E}}(y):=\sin{(\theta_{y, \mathrm{E}}(y))}, \\[2pt]
P_x:=\dyad{x-a, y}{x, y},\quad
&Q_x:=\dyad{x+a, y}{x, y}, \\[2pt]
P_y:=\frac{1}{2}(\dyad{x, y-a}{x, y}+\dyad{x, y+a}{x, y}), \quad%
&Q_y:=\frac{1}{2}(\dyad{x, y-a}{x, y}-\dyad{x, y+a}{x, y}). \label{eqQABP213}
\end{align}
We let $\{\sigma^x, \sigma^y, \sigma^z\}$, $\{\tau^x, \tau^y, \tau^z\}$ and $\{\upsilon^x, \upsilon^y, \upsilon^z\}$ denote the Pauli matrices for the spaces spanned by $\{\lk, \rk\}$, $\{\dk, \uk\}$ and $\{\ket{\mathrm{G}}, \ket{\mathrm{E}}\}$, respectively.
The identity matrix for each space is given by $\sigma^0$, $\tau^0$ and $\upsilon^0$, respectively, and $\upsilon^{\mathrm{G/E}}:=(1\pm\upsilon^z)/2$.
Note that $S_x^\textrm{2d}$ and $S_y^\textrm{2d}$ are identical to the ones in equation~(17) in Ref.~\cite{Yamagishi23} except for the factor $\upsilon^0$, while $C_x^\textrm{2d}$ and $C_y^\textrm{2d}$ are extensions of the ones in equation~(17) in Ref.~\cite{Yamagishi23} with the factors $\upsilon^{\mathrm{G}}$ and $\upsilon^{\mathrm{E}}$.

\subsection{Numerical results for the two-dimensional model}

\begin{figure}
  \centering
  \vspace{-0.4\baselineskip}
    \includegraphics[width=0.4\columnwidth]{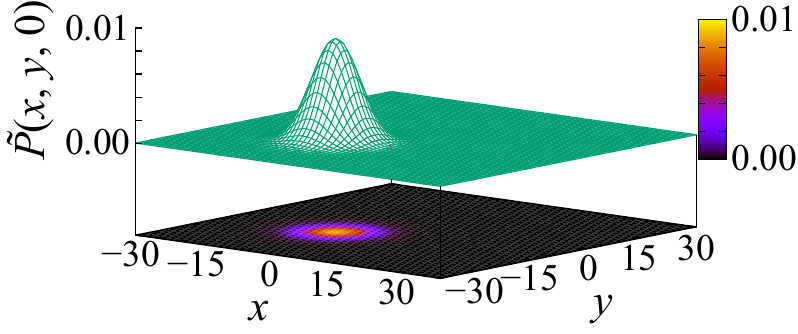}
    \caption{An eigenstate for the eigenvalue close to unity, $U_n=0.9336010518344118+\ii\,0.3583142140826795$, with the parameters in equation~\eqref{eqQABP27} and $g=0$, which was numerically computed. The eigenstate is shifted $(\delta_x, \delta_y)=(9, 0)$ steps and is induced initial velocity in the form of $\ee^{\ii(k_xx+k_yy)}$ with $(k_x, k_y)=(0, \pi)$ to be made into the initial sate for the computation of dynamics.}
    \label{qabp_29_kugui_init}
\end{figure}

For numerical calculation in two dimensions, we set the system size to $L_x=L_y=71$ with $-35\le x\le35$ and  $-35\le y\le35$.
We again introduce linear potentials as phases of the coin operators:
\begin{align}
&\theta_{x, \mathrm{G/E}}(x)=\begin{cases}
\theta_{\mathrm{g/e}} & \mbox{for $x<\frac{-\alpha-1}{\beta}$} \\
\theta_{\mathrm{g/e}}(\alpha+\beta x) & \mbox{for $\frac{-\alpha-1}{\beta}\le x\le\frac{-\alpha+1}{\beta}$} \\
-\theta_{\mathrm{g/e}} & \mbox{for $x>\frac{-\alpha+1}{\beta}$}
\end{cases}, \label{eqQABP25} \\
& \theta_{y, \mathrm{G/E}}(y)=\begin{cases}
\theta_{\mathrm{g/e}} & \mbox{for $y<\frac{-\alpha-1}{\beta}$} \\
\theta_{\mathrm{g/e}}(\alpha+\beta y) & \mbox{for $\frac{-\alpha-1}{\beta}\le y\le\frac{-\alpha+1}{\beta}$} \\
-\theta_{\mathrm{g/e}} & \mbox{for $y>\frac{-\alpha+1}{\beta}$}
\end{cases}, \label{eqQABP26} \\
&\theta_{\mathrm{g/e}}=\theta_0\pm\varepsilon.
\end{align}
The potential of equation~\eqref{eqQABP25} and \eqref{eqQABP26} 
squared is a  harmonic potential~\cite{Yamagishi23} with its top cut off.

All the computation described throughout this section was conducted with the parameters fixed as follows:
\begin{align}\label{eqQABP27}
\theta_0=\frac{\pi}{8},\quad\varepsilon=0.25,\quad w=0.25,\quad\alpha=0.5,\quad\beta=0.05.
\end{align}
\noindent
For the initial state, we used an eigenstate of the eigenvalue close to unity (Fig.~\ref{qabp_29_kugui_init}) shift to $x$ by nine sites and imposed the initial velocity in the form of $\ee^{\ii(k_xx+k_yy)}$ with $(k_x, k_y)=(0, \pi)$.

\begin{figure}
  \centering
  \includegraphics[width=0.5\columnwidth]{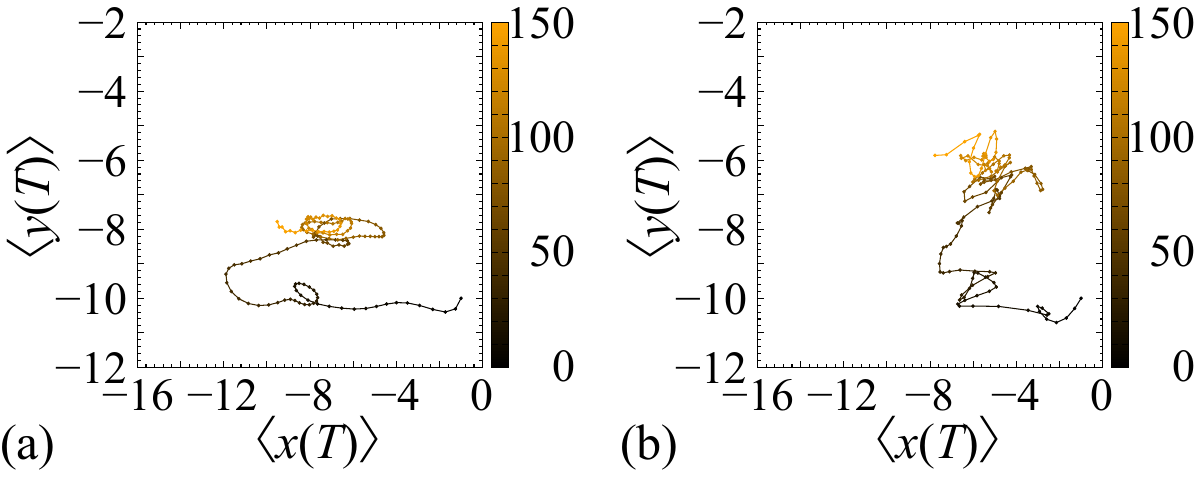}
  \caption{The orbit of the quantum walker for (a) $g=0$ and (b) $g=1$. The vertical axis indicates $\expval{y(T)}$ and the horizontal axis indicates $\expval{x(T)}$. Black circles indicate the values at the beginning of time evolution and they turn into orange as time goes on.}
  \label{qabp_29_kugui_orbit}
\end{figure}

Figure~\ref{qabp_29_kugui_orbit} shows the expectation values,
\begin{align}
\expval{x(T)}:=\sum_{x, y}{x \tilde{P}(x, y, T)},\quad
\expval{y(T)}:=\sum_{x, y}{y \tilde{P}(x, y, T)} \label{eqQABP28}
\end{align}
at each time step, where
\begin{align}
P(x,y,T):=\abs{\braket{x,y}{\psiR(T)}}^2,\quad
\tilde{P}(x, y, T)%
:=\frac{P(x,y,T)}{\sum_{x,y}P(x,y,T)}.
\end{align}
We observe that our quantum active particle moves around in wider region with larger values of $g$.
This is a result similar to the previous research~\cite{Schweitzer98} in a classical system; see also Fig.~\ref{qabp_29_kugui_ani_40}.
Meanwhile, we observe in Fig.~\ref{qabp_29_kugui_orbit}(a) that the movement of our quantum active particle almost converges to a limit cycle for $g=0$, not falling into the origin.
This is a quantum feature: our quantum active particle stays on the constant energy surface.

\begin{figure}
  \centering
  \includegraphics[width=0.5\columnwidth]{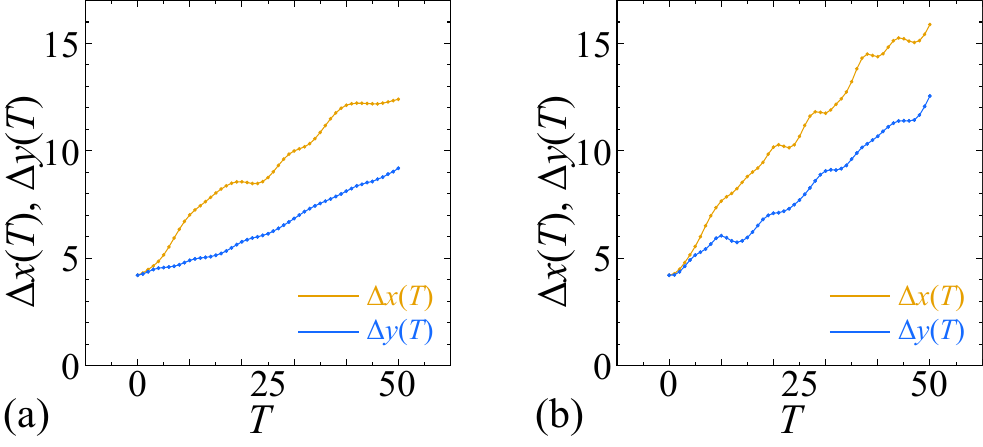}
  \caption{The time-step dependence of the standard deviations $\Delta x$ (yellow lines) and $\Delta y$ (blue lines) for (a) $g=0$ and (b) $g=1$. We only plot standard deviations during $0\le T\le50$ since standard deviations outside the plotting range are not reliable due to the periodic boundary condition; wave functions reach the edge by around $T\simeq50$ as you can see in Fig.~\ref{qabp_29_kugui_ani_40}.}
  \label{qabp_29_kugui_SD}
\end{figure}

\begin{figure}
  \centering
  \includegraphics[width=0.8\columnwidth]{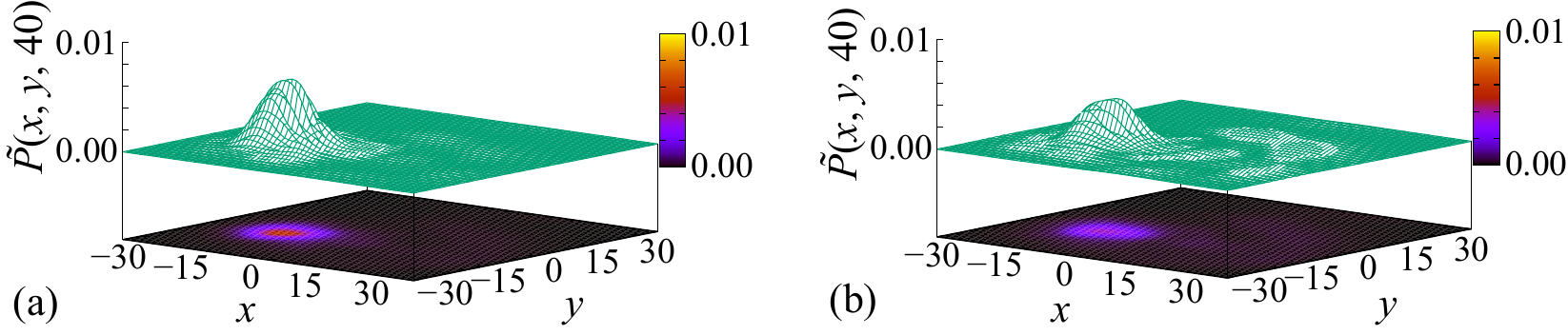}
  \caption{The normalized probability distribution of the quantum walker $\tilde{P}(x,y,40)$ after $40$ time steps of time evolution for (a) $g=0$ and (b) $g=1$.}
  \label{qabp_29_kugui_ani_40}
\end{figure}

The standard deviations $\Delta x(T)$ and $\Delta y(T)$ of the walker at each time step
\begin{align}
\Delta x(T):=\sqrt{\sum_{x, y}{\mqty[(x-\expval{x(T)})^2 \tilde{P}(x, y, T)]}},\quad
\Delta y(T):=\sqrt{\sum_{x, y}{\mqty[(y-\expval{y(T)})^2 \tilde{P}(x, y, T)]}} \label{eqQABP211}
\end{align}
are shown in Fig.~\ref{qabp_29_kugui_SD}.
We can clearly see that the standard deviation tends to take larger values as $g$ increases.
This means that our quantum active walker moves around more actively when it takes up more energy from the environment.
This is also a result similar to the previous research~\cite{Schweitzer98} in a classical system.
Meanwhile, we again observe oscillation in Fig.~\ref{qabp_29_kugui_SD} as in Fig.~4 in the main text
This resonant transition between two states is again one of the quantum features that we observe.

The normalized probability distribution of the sum of the ground and the excited states at each site after 40 time steps are shown in Fig.~\ref{qabp_29_kugui_ani_40}.
We can see that the side peaks 
become relatively larger compared to the peak around the center as $g$ increases.


\bibliographystyle{hunsrt}
\bibliography{yamagishi}